\documentclass[preprint,10pt]{elsarticle}

\usepackage[utf8]{inputenc}
\usepackage[english]{babel}
\usepackage{amssymb}
\usepackage[utf8]{inputenc}
\usepackage[english]{babel}
\usepackage{natbib}
\usepackage{hyperref}
\usepackage{amsmath}
\usepackage{longtable}
\usepackage{geometry}

\biboptions{sort&compress}
\journal{Nuclear Physics A}

\begin{document}

\begin{frontmatter}



\title{Superheavy Magic Nuclei: Ground-State Properties, Bubble Structure and $\alpha$-Decay Chains}


\author[a]{R. Sharma}\author[a,b]{A. Jain}\author[c]{M. Kumawat}\author[d]{J. K. Deegwal}\author[e]{Abdul Quddus}\author[f]{G. Saxena}
\address[a]{Department of Physics, S. S. Jain Subodh P.G.(Autonomous) College, Jaipur-302004, India}
\address[b]{Department of Physics, School of Basic Sciences, Manipal University Jaipur, Jaipur-303007, India}

\address[c]{Department of Science and Technology, Faculty of Education and Methodology, Jayoti Vidyapeeth Women's University, Jaipur-303007, India}
\address[d]{Govt. Women Engineering College, Ajmer-305002, India}
\address[e]{Applied Sciences and Humanities Section, University Polytechnic, Aligarh Muslim University, Aligarh - 202002, UP, India.}
\address[f]{Department of Physics (H\&S), Govt. Women Engineering College, Ajmer-305002, India}
\begin{abstract}
A systematic investigation of superheavy nuclei in the isotopic chains of proton numbers Z$=$106, 114, 120, and 126 together with isotonic chains of neutron numbers N$=$162, 172, and 184 is presented in the theoretical framework of relativistic mean-field density functionals based on density-dependent meson-nucleon couplings. Ground-state properties, including binding energy, shape, deformation, density profile, and radius, are estimated to provide compelling evidence of magicity in these even-even nuclei, aligning with the concept of the 'island of stability.' The analysis reveals central depletion in the charge density, indicating a bubble-like structure, primarily attributed to the substantial repulsive Coulomb field and the influence of higher $l$-states. A thorough examination of potential decay modes, employing various semi-empirical formulas, is presented. The probable $\alpha$-decay chains are evaluated, demonstrating excellent agreement with available experimental data.
\end{abstract}




\end{frontmatter}


\section{Introduction}
In the past four decades, there has been significant progress in understanding the nuclear shell structure, spanning from light nuclei to superheavy nuclei (SHN). This advancement has led to the identification of new magic numbers and the subsequent conceptualization of the 'island of stability.' The journey undertaken by experimentalists in the pursuit of SHN is marked by numerous experiments conducted at GSI, Darmstadt  \cite{hofmann2000,Hamilton2013}, RIKEN, Japan \cite{Morita2004}, JINR-FLNR, Dubna \cite{Oganessian2004,Oganessian2015}, and Argonne National Laboratory in USA \cite{Davids1989}, etc.
These experiments have not only unveiled the existence of new heavy isotopes of superheavy elements but have also validated the concept of the island of enhanced stability near N$\sim$184 \cite{ogan2011}. This stability is a characteristic signature of magicity, anticipated in the superheavy region, mirroring the established pattern in the rest of the periodic chart.\par
There are several observations in the superheavy region that offer evidence of shell gaps and demonstrate stability in $^{268}$Hs (N$=$160) \cite{nishio2010}, $^{270}$Hs (N$=$162) \cite{dvorak2006,oganessian2013}, as well as at N$=$152 with Z$=$98-108 \cite{oganessian2017}. Theoretical investigations of the shell structure of SHN using various approaches reveal that magicity beyond the conventional spherical doubly magic nucleus $^{208}$Pb (Z$=$82; N$=$126) is somewhat model-dependent. Microscopic calculations predict various stability regions, such as Z$=$114; N$=$184 \cite{Li2017,Manju2020,Rutz1997,Seyyedi2a0,Zhang2012}, Z$=$120; N$=$172 \cite{Rutz1997,Zhang2005,Seyyedi2a0,Patra2014,Ismail2017,Beckmann2000,Patra2021,Patra2012}, Z$=$120; N$=$184 \cite{Ghodsi2020,Zhang2005,Siddiqui2020}, and Z$=$124 or 126; N$=$184 \cite{Siddiqui21, Manju2020, Rutz1997, Kruppa00, Cwiok96, Cwiok05}. Recently, from relativistic mean-field theory, the shell gap responsible for the stability has been reported at Z$=$120; N$=$164 \cite{saxena1} along with Z$=$104; N$=$184, Z$=$102; N$=$152, Z=108; N$=$162 \cite{singhnpa2020}, etc. However, an exhaustive and in-depth study of all these predicted stability regions is required, representing the primary objective of this paper.\par

A feature that has recently garnered significant attention in magic nuclei of the lighter mass region and has found very promising in the superheavy region is the phenomenon known as 'central depletion' or 'bubble structure.' Based on contemporary studies focusing on the mechanism of bubble structure \cite{li,schuetrumpf,decharge,sksingh,berger2001,afanasjev2005}, the primary reason for the reduction in the charge density at the nucleus's center in the heavy or superheavy region can be attributed to either (i) empty s-states, as investigated theoretically \cite{li,schuetrumpf} and experimentally verified in sd-shell nuclei \cite{nature}, or (ii) a substantial repulsive electrostatic Coulomb field, which predominantly occurs in SHN due to the large number of protons. This phenomenon has been explored using relativistic Hartree-Fock-Bogoliubov (RHFB) theory to identify the bubble-like structure in nuclei with Z$=$120 \cite{li}. Furthermore, correlation analysis was conducted for isotonic chains of N$=$82, 126, and 184, applying Skyrme functionals in nuclear density functional theory (nuclear-DFT) \cite{schuetrumpf}, supporting the dominance of the Coulomb effect on the bubble structure. Additionally, the prediction of the occurrence of the bubble-like structure in superheavy magic numbers (N$=$164, 184, 228) and in a few isotopes of Z$=$122, 120, 118 has been made in some of our recent works \cite{saxenaijmpe2018,saxenaijmpe2019}. This phenomenon is also anticipated in potential superheavy magic candidates analogous to the light mass region \cite{saxena}, constituting one of the motives for the present study.\par

To date, the heaviest synthesized element is $^{294}$118 \cite{ogan2006}. For the production of elements beyond Oganesson (Z$=$118), complete fusion reactions with projectiles having Z$>$20 are required, primarily due to the limited availability of actinide targets for Z$>$98 \cite{hofmann2016,Adamian2004}. Both theoretical and experimental approaches have been employed to produce elements with Z$>$118, such as $^{50}$Ti + $^{249}$Bk for elements Z$=$119 and Z$=$120 \cite{Adamian2020}, along with $^{50}$Ti + $^{249}$Cf for the production of the element with Z$=$120 \cite{Albers2020}, among others. Almost all such experiments rely on $\alpha$-decay, a crucial decay process configuration that has proven beneficial for the production and investigation of SHN. While spontaneous fission (SF) is equally important to $\alpha$-decay, as it aids in a similar manner in preparing novel nuclei in the forecasted magic island around the SHN Z$=$114 to Z$=$126 due to the shell effect. Therefore, theoretical investigations are vital as they can provide essential inputs related to the decay properties of SHN, including $\alpha$-decay \cite{Qi2016,Delion2015,Poenaru2006,Chowdhury2008,Denisov2009,saxenaPSS2021,Singh2020,Saxena2023epja} and SF \cite{zagrebaev2012,Möller2019,Heßberger2016,karpov2012,sarriguren2019}. Theoretical probes serve as a platform for experimentalists to design and plan their experiments.\par

Our current study is structured in three steps: (i) the evaluation of shapes and various ground-state properties of SHN to establish their magic character in selected cases, (ii) a systematic investigation of central depletion in connection with magicity in the considered SHN, and (iii) the examination of possible decay modes, namely $\alpha$-decay and SF, leading to potential $\alpha$-decay chains in these identified magic nuclei. Our calculations for SHN yield significant insights into their ground-state properties, magicity, the existence of bubble structures, and potential decay modes.
\label{intro}
\section{Formalism and Calculations}
\subsection{Relativistic Mean-Field (RMF) Theory}
Our theoretical calculations are carried out with the help of the following Lagrangian density
\begin{eqnarray}
{\cal L}& = &{\bar\psi}(\imath \gamma.\partial- m)\psi + \frac{1}{2}(\partial\sigma)^{2}- \frac{1}{2}m_{\sigma}\sigma^2 -\frac{1}{4}H_{\mu \nu}H^{\mu \nu}+ \frac{1}{2}m_{\omega}^{2}\omega^{2} -\frac{1}{4}\overrightarrow{R}_{\mu \nu}\overrightarrow{R}^{\mu \nu}
+ \frac{1}{2}m_{\rho}^{2}\overrightarrow{\rho}^{2}\nonumber \\
&&-\frac{1}{4}F_{\mu \nu}F^{\mu \nu}-g_{\sigma}{\bar\psi} \sigma \psi
- g_{\omega}{\bar\psi} \gamma.\omega\psi- g_{\rho}{\bar\psi} \gamma.\overrightarrow{\rho}\tau\psi- e{\bar\psi} \gamma.A\frac{(1-\tau_{3})}{2} \psi.
\end{eqnarray}
where, $H$, $G$ and $F$ are the field tensor for the vector fields as defined by
\begin{eqnarray}
                 H^{\mu \nu} &=& \partial^{\mu} \omega^{\nu} -
                       \partial^{\nu} \omega^{\mu}\nonumber\\
                 \overrightarrow{R}^{\mu \nu} &=& \partial^{\mu} \overrightarrow{\rho}^{\nu}-\partial^{\nu} \overrightarrow{\rho}^{\mu}\nonumber\\
                  F^{\mu \nu} &=& \partial^{\mu} A^{\nu} -
                       \partial^{\nu} A^{\mu}\,\,\nonumber\
\end{eqnarray}
The symbols used here retain their conventional meanings. The Lagrangian includes the standard Yukawa coupling between the nucleonic field ($\psi$) and various mesonic fields, specifically the isoscalar-scalar $\sigma$, isoscalar-vector $\omega$, isovector-vector $\rho$, and photon $A$. The couplings of nucleons with the $\sigma$, $\omega$, and $\rho$ mesons are represented by $g_{\sigma}$, $g_{\omega}$, and $g_{\rho}$, respectively. These couplings exhibit an explicit density dependence formulated as follows \cite{Lalazissis05}:

\begin{equation} \label{eq:Equation1} 
g_{i}(\rho) =  g_{i}(\rho_{sat})f_{i}(x), \,\,\,\,\,\,\, \text{for} \,\,\, i = \sigma, \omega 
\end{equation}

where the density dependence is given by
\begin{equation} \label{eq:Equation2} 
f_{i}(x) =  a_{i} \frac {1+b_{i}(x+d_{i})^{2}}{1+c_{i}(x+e_{i})^{2}} 
\end{equation}
in this expression, $x$ is defined as $x = \rho/\rho_{\text{sat}}$, where $\rho_{\text{sat}}$ represents the baryon density at saturation in symmetric nuclear matter. The density dependence for the $\rho$ meson takes an exponential form and is expressed as follows:
\begin{equation} \label{eq:Equation3} 
f_{\rho}(x) =  exp(-a_{\rho}(x-1)) 
\end{equation}
This model is referred to as the density-dependent meson-exchange model (DD-ME), and our calculations utilize the DD-ME2 parameter set. A separable form of the Gogny force in the pairing channel has been used in the Relativistic Hartree Bogoliubov calculations  \cite{Niksic10}. For a more in-depth understanding of these formulations, we direct the reader to Ref.~\cite{Lalazissis05}.

\subsection{$\alpha$-Decay}
\label{decay}
To compute $\alpha$-decay half-lives, one needs the disintegration energy $Q_\alpha$ for ground-state to ground-state decay. This value can be derived from mass excesses or total binding energies, incorporating the calculated energy release using the following equation:
\begin{eqnarray}
         Q_\alpha(Z, N) & = & M(Z, N) - M(Z-2, N-2)-  M(2, 2) \nonumber\\
                         & =& B.E.(Z-2, N-2) + B.E.(2, 2) - B.E.(Z, N)
\label{qalpha}
\end{eqnarray}
In this context, the mass excess of $^{4}$He, denoted as M(2,2), is 2.42 MeV, and the binding energy B.E.(2,2) is 28.30 MeV. To calculate $\alpha$-decay half-lives ($log_{10}T^{\alpha}_{1/2}$), we utilize several recently fitted and modified empirical formulas \cite{singhnpa2020,saxena2020prc}, including formulas that consider the magicity of the nucleus. The ensuing section provides a brief description of these empirical formulas.

\subsubsection{Modified Manjunatha formula (MMF)}
The initial formula utilized in this context is the modified version of the Manjunatha formula (MMF) \cite{singhnpa2020}. It incorporates isospin dependence and features a second-order polynomial term of $Z_d^{0.4}/\sqrt{Q_{\alpha}}$, expressed as follows:

\begin{eqnarray}
 log_{10}T^{\alpha}_{1/2}(s) &=&  a (Z_d^{0.4}/\sqrt{Q_{\alpha}})^2 + b (Z_d^{0.4}/\sqrt{Q_{\alpha}}) + c + dI + eI^2
 \label{modify-alpha}
\end{eqnarray}

Here, $Z_d$ denotes the atomic number of the daughter nucleus, and $I = \frac{N-Z}{A}$ represents the isospin. The coefficients employed in this formula are: $a=-5.5733$, $b=48.1849$, $c=-80.3574$, $d=5.1224$, and $e=27.4312$ for even-even nuclei.

\subsubsection{Modified Horoi formula (MHF)}
The second formula for $\alpha$-decay is the modified version of the scaling law for cluster decay, known as the modified Horoi formula (MHF) \cite{saxena2020prc}. It is expressed as:

\begin{eqnarray}\label{horoi-modi}
log_{10}T^{\alpha}_{1/2}(s) &=& (a \sqrt{\mu} + b)[(Z_c Z_d )^{0.6} Q_{\alpha}^{-1/2} - 7]+ (c \sqrt{\mu} + d) + eI + fI^2
\end{eqnarray}

Here, $\mu = \frac{A_d A_c}{A_d + A_c}$ represents the reduced mass, $I$ is the isospin, and the coefficients for even-even nuclei are: $a=834.3669$, $b=-1648.1873$, $c=-875.9054$, $d=1722.1682$, $e=90.1158$, and $f=-267.2140$.

\subsubsection{Modified Sobiczewski formula (MSF)}
The third formula considered for the present study is the latest fitted and modified version of the widely used phenomenological Sobiczewski formula (MSF) \cite{saxena2020prc}, represented by the following equation:
\begin{equation}\label{sobic-modi}
log_{10}T^{\alpha}_{1/2}(s) = aZ(Q_{\alpha} - \overline{E}_i)^{-1/2} + bZ + c + dI +eI^2
\end{equation}
For the present study, only even-even nuclei are considered, hence the value of $\overline{E}_{i}$ is taken as zero. The coefficients for the formula are: $a=1.2659$, $b=-0.1439$, $c=-23.3040$, $d=-81.2157$, and $e=275.6497$.

\subsubsection{SemFIS2}Another semiempirical formula employed in the present study is based on the fission theory of $\alpha$ decay, considering the magic numbers of nucleons. This formula, known as SemFIS \cite{poenaru2006}, has been subsequently modified as SemFIS2 \cite{poenaru2007}, and is expressed as:

\begin{equation}
log_{10}T^{\alpha}_{1/2}(s) = 0.43429K_{s}\chi(x,y)-20.446+H^f
\end{equation}
where
\begin{eqnarray}
K_{s} &=& 2.52956Z_d[A_d/(AQ_{\alpha})]^{1/2} [arccos \sqrt{r}-\sqrt{r(1-r)}] \nonumber
\end{eqnarray}
and
\begin{equation}
r = 0.423Q_{\alpha}(1.5874 + A_d^{1/3})/Z_d \nonumber
\end{equation}
where $\chi(x,y)$ is the functional-coefficient given as:
\begin{equation}
\chi(x,y)=B_1 +x(B_2 + xB_4 )+y(B_3 + yB_6 ) +xyB_5 \nonumber
\end{equation}
The coefficients are: B$_{1}$=0.9854, B$_{2}$=0.1022, B$_{3}$=-0.0249, B$_{4}$=-0.8321, B$_{5}$=1.5057, and B$_6$=-0.6812. The reduced variables $x$, $y$ are defined as: $x$=(N-127)/(185-127), and, $y$=(Z-83)/(127-83). The value of $H^{f}$ is taken zero for even-even nuclei.

\subsubsection{The Universal Curve (UNIV)}
Poenaru \textit{et al.} \cite{poenaru2011} derived the single line of the universal (UNIV) curve for $\alpha$ and cluster radioactivities by plotting the sum of the decimal logarithm of the half-life and the cluster preformation probability as a function of the decimal logarithm of the penetrability of the external barrier. This formula is known as the UNIV formula for $\alpha$-decay half-lives and can be described as:
\begin{eqnarray}
log_{10}T^{\alpha}_{1/2}(s) &=& -log_{10}P-log_{10}S_{\alpha} +[log_{10}(ln2)-log_{10}\upsilon]
\end{eqnarray}

The penetrability of an external Coulomb barrier P may be obtained analytically as
\begin{eqnarray}
-log_{10}P &=& 0.2287\sqrt{\mu Z_d Z_{\alpha}R_b} \times[arccos\sqrt{r}-\sqrt{r(1-r)}] \nonumber
\end{eqnarray}
where
$r$=$R_{a}/R_{b}$ with $R_a (fm)=1.2249(A_{d}^{1/3}+A_{\alpha}^{1/3}$), $R_b(fm)=1.4399Z_d Z_{\alpha}/Q_{\alpha}$ being the two classical turning points. The term
$log_{10}S_{\alpha}$ is the logarithm of preformation factor which is given by
\begin{equation}
log_{10}S_{\alpha} = -0.598 (A_{\alpha}-1)
\end{equation}
and the last term $[log_{10}(ln2)-log_{10}\upsilon]$ is an additive constant with the value -22.1692.

\subsection{Spontaneous Fission}
 For a limited number of heavy isotopes, another decay mode, i.e., spontaneous fission (SF), is strong enough to compete with the $\alpha$-decay mode. The SF half-life was first proposed by Swiatecki \cite{wjswiatecki1955} based on the fission barrier heights and the values of the fissility parameter $Z^2/A$. Subsequently, many other attempts \cite{dwdorn1961,cxu2005,karpov2012} have been made to improve the formula. In our present work, the SF half-lives of the considered nuclei have been evaluated using the modified Bao formula (MBF-2020) \cite{saxena2020prc}, which is given by:

\begin{eqnarray}\label{baoSF}
log_{10}T_{1/2}^{SF} (s) &=& c_1 + c_2 \left(\frac{Z^2}{(1-kI^2)A}\right)   + c_3 \left(\frac{Z^2}{(1-kI^2)A}\right)^2 + c_4 E_{s+p}
\end{eqnarray}

here, k$=$2.6 and other coefficients are: c$_1$$=$893.2644 (for even-even nuclei), c$_2$$=$$-$37.0510, c$_3$$=$0.3740, c$_4$$=$3.1105.

\subsection{Uncertainty Analysis}
Superheavy nuclei exhibit significantly higher experimental uncertainties in measured half-lives compared to the rest of the nuclear chart, often by an order of magnitude. Therefore, accounting for these uncertainties when predicting nuclear properties is crucial for ensuring the accuracy and completeness of the results. According to NUBASE2020 \cite{kondev2021}, there are approximately 160 experimental data points for $Q$ values and 56 data points for half-lives for nuclei with Z$>$104. The uncertainty in these $Q$ values ranges from 0.004 MeV to 0.99 MeV, with an average uncertainty of $\pm$0.21 MeV. Similarly, the uncertainty in the available experimental half-lives varies from
10$^{-6}$ s to 198 s with an average value $\pm$8.01 s.\par

Given that the accuracy of formulas predicting $\alpha$-decay half-lives is heavily dependent on the precision of $Q$ values, incorporating these uncertainties is essential for meaningful comparisons with experimental data. Our analysis reveals that the uncertainty in theoretical $Q$ values, when compared to 180 experimental data points, is $\pm$0.04 MeV.  This statistical uncertainty is calculated using the following formula in our obtained $Q$ values from DD-ME2 parameter.

\begin{equation}\label{uncertainty}
  u = \sqrt{\frac{\sum(x_i - \mu)^2}{n(n-1)}}
\end{equation}
where, $x_i$ is the $i^{th}$ reading in the data set, $\mu$ is the mean of the data set and n is the number of readings in the data set. By incorporating these uncertainties into the various considered formulas, we have determined the average uncertainties in the predicted logarithmic half-lives, as shown in Table \ref{uncertainty}. Notably, all theoretical predictions for $\alpha$-decay half-lives presented in this work fall within the stated uncertainty range, reinforcing the robustness and reliability of the current analysis.

\begin{table}[h!]
\centering
\caption{Theoretical uncertainties in the $log_{10}T_{1/2}$ values (s). obtained in several considered empirical formulas.}
\label{uncertainty}
\begin{tabular}{|c@{\hskip 0.5in}|@{\hskip 0.3in}c|}
\hline
\textbf{Formula} & \textbf{Uncertainty} \\ \hline
 MMF              & $\pm$0.09                  \\ \hline
 MSF               &$\pm$0.18                  \\ \hline
 MHF               &$\pm$0.30                  \\ \hline

 SemFIS2               &$\pm$0.15                  \\ \hline
 UNIV               &$\pm$0.22                  \\ \hline

\end{tabular}
\end{table}
\par

\section{Results and discussions}
For superheavy nuclei, the location of the valley of $\beta$-stability is predicted to shift significantly towards neutron-rich isotopes due to the increased Coulomb repulsion from the high number of protons. Theoretical models suggest that the valley of $\beta$-stability  for these elements resides around proton numbers Z$\approx$114 to Z$\approx$126 and neutron numbers N$\approx$184. Within this region, known as the "island of stability," nuclei are expected to exhibit enhanced stability due to large shell gaps and closed proton and neutron shells. Consequently, while superheavy elements are predominantly unstable with short half-lives due to decay modes like $\alpha$-decay and spontaneous fission, those within this predicted island of stability are anticipated to have significantly longer half-lives compared to their neighbors. Despite experimental challenges, ongoing research aims to explore this region to confirm the existence of such stable or semi-stable superheavy nuclei and refine our understanding of nuclear stability at the extremes of the periodic table.\par

Over the past several decades, numerous theoretical investigations have explored the possibility of extra stability for various neutron and proton numbers in the superheavy region. As mentioned previously, the existence of an island of stability has been suggested, particularly around N$\sim$184 in certain experiments. In this context, we have compiled an extensive table of probable superheavy magic numbers, along with their respective references, to identify common and authoritative magic numbers for our study. Table \ref{magic} of the appendix presents proton and neutron magic numbers (first column) alongside their corresponding neutron/proton numbers (second column), where they are expected to exhibit a magic character or a relatively stable configuration.\par
From Table \ref{magic} of the appendix, we have specifically chosen proton magic numbers Z$=$106, 114, 120, 126, and neutron magic numbers N$=$162, 172, and 184 to assess their credibility as magic numbers and investigate their structural properties. This analysis is conducted using the RMF approach with the DD-ME2 force parameter in the first subsection. Additionally, we explore central depletion in the charge density of SHN associated with these magic numbers. In the second subsection, we delve into the competition between probable decay modes ($\alpha$-decay and SF) in the full isotopic chains of Z$=$106, 114, 120, 126. Consequently, we report possible $\alpha$-decay chains along with their estimated half-lives.

\subsection{Ground-State Properties}
\begin{figure*}[!htbp]
\centering
\includegraphics[width=1.0\textwidth]{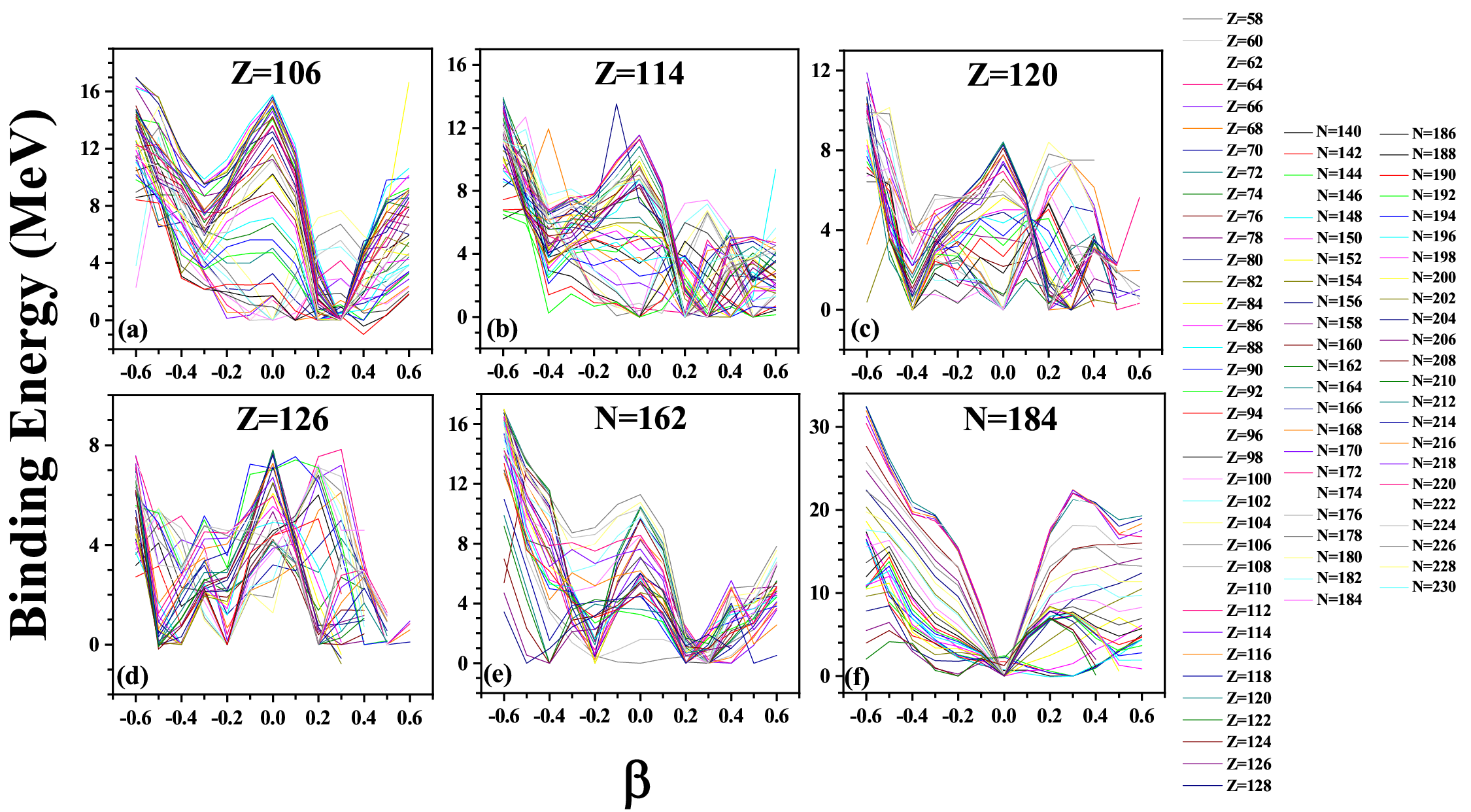}
\caption{(Colour online) Potential energy surfaces for nuclei with Z$=$106, 114, 120, 126 isotopes and N$=$162 and 184 isotones obtained from DD-ME2 \cite{Lalazissis05} parameter.}\label{pes}
\end{figure*}
As previously mentioned, quadrupole-constrained calculations are performed using the DD-ME2 parameter for even-even nuclei of all selected isotopes and isotones up to N$\leq$220 and Z$\leq$128, respectively. The corresponding potential energy surfaces (PESs) are obtained and depicted in Fig. \ref{pes} as a function of $\beta$ (quadrupole deformation parameter). The binding energies are normalized to zero corresponding to the lowest values of energy obtained for each isotope of Z$=$106, 114, 120, 126, and similarly for isotones of N$=$162 and 184. This analysis aims to understand their shapes and the value of ground-state deformation.\par

Figs. \ref{pes}(a), (b), and (c) reveal a diverse range of shapes for the considered nuclei, with the majority found to be well-deformed. For Z$=$106, 114, and Z$=$120 isotopes, most nuclei exhibit dominant prolate shapes, while many of the Z$=$126 (Fig. \ref{pes}(d)) nuclei possess oblate shapes. Interestingly, a few nuclei are found to have shapes coexisting with both oblate and prolate nature. A similar coexistence phenomenon of oblate and prolate shapes is observed for the isotonic chain of N$=$162. Additionally, several cases of spherical shapes as dominant shapes are reported in Fig. \ref{pes} for the N$=$184 isotones. The region around N$\sim$184 is anticipated as the 'island of stability,' as evidenced by the discovery of the heaviest synthesized element $^{294}$118 \cite{ogan2006} and eleven new heaviest isotopes of elements Z$=$105 to Z$=$117 \cite{ogan2011} through the $\alpha$-decay chains in this region of enhanced stability (near N$\sim$184).\par

It is well-established that $\alpha$-decay is the most probable decay mode for heavy and SHN, facilitating their detection in laboratory experiments. The $\alpha$-decay energies (Q${\alpha}$) are known for several SHN, making them valuable for validating theoretical predictions. Moreover, a lower Q${\alpha}$ value compared to neighboring nuclei indicates the stability of the nucleus, a crucial aspect in the study of magic nuclei.\par

In this context, Fig. \ref{qalpha} displays Q$_{\alpha}$ values calculated using the DD-ME2 parameter for the considered proton and neutron magic nuclei. The comparison with available experimental data \cite{nndc} confirms the validity of the RMF theory in this region of the periodic chart. This validation serves as a foundation for predicting $\alpha$-decay chains in the subsequent subsections of this article. Fig. \ref{qalpha} affirms the stability of neutron numbers N$=$162, 172, 184, and 196, in line with the findings presented in Table \ref{magic} of the appendix.\par

\begin{figure*}[!htbp]
\centering
\includegraphics[width=0.9\textwidth]{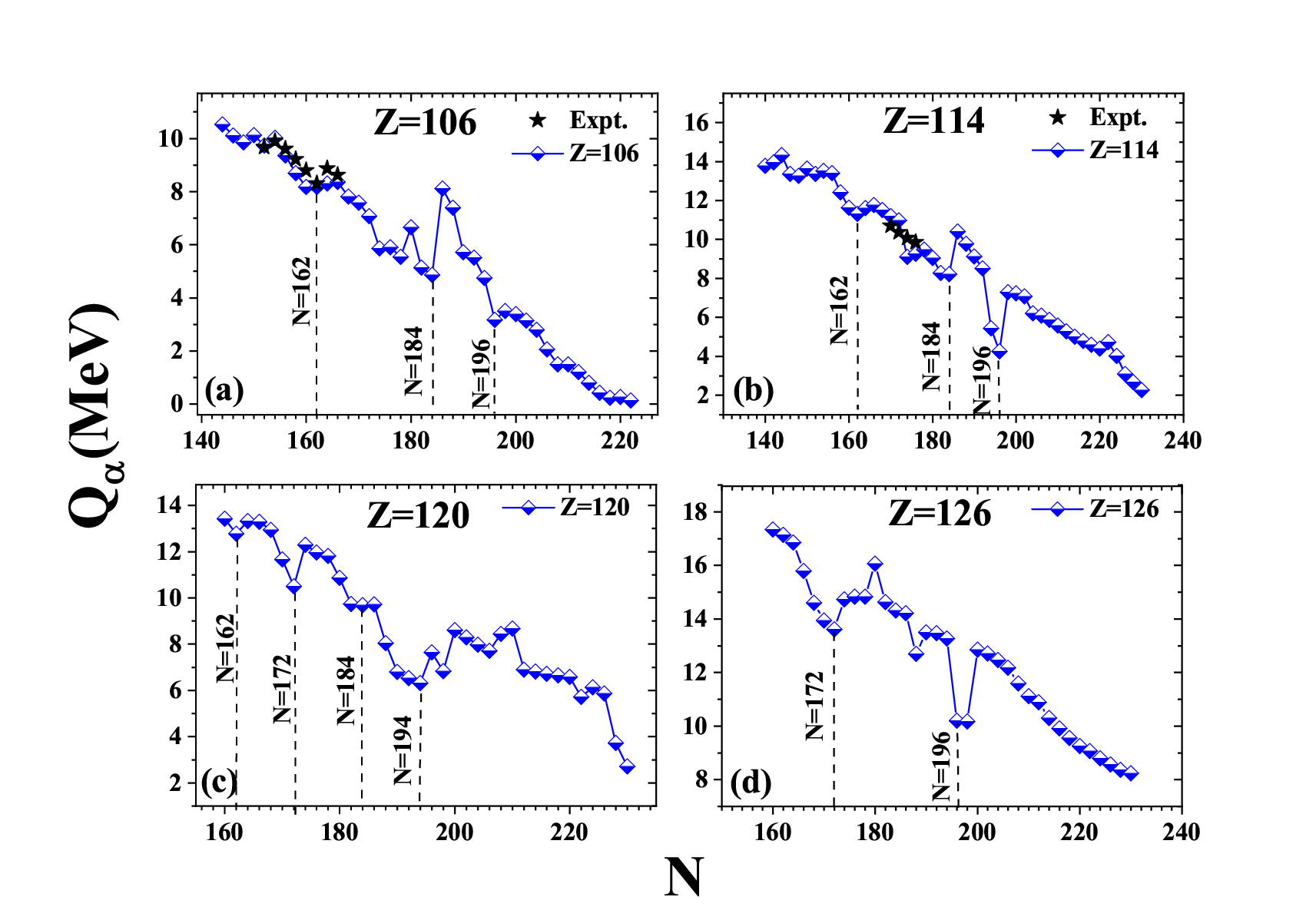}
\caption{(Colour online) $\alpha$ - decay energies for nuclei with Z$=$106, 114, 120, and 126 obtained from RMF approach using DD-ME2 \cite{Lalazissis05} force parameter.}\label{qalpha}
\end{figure*}
\begin{figure*}[!htbp]
\centering
\includegraphics[width=0.7\textwidth]{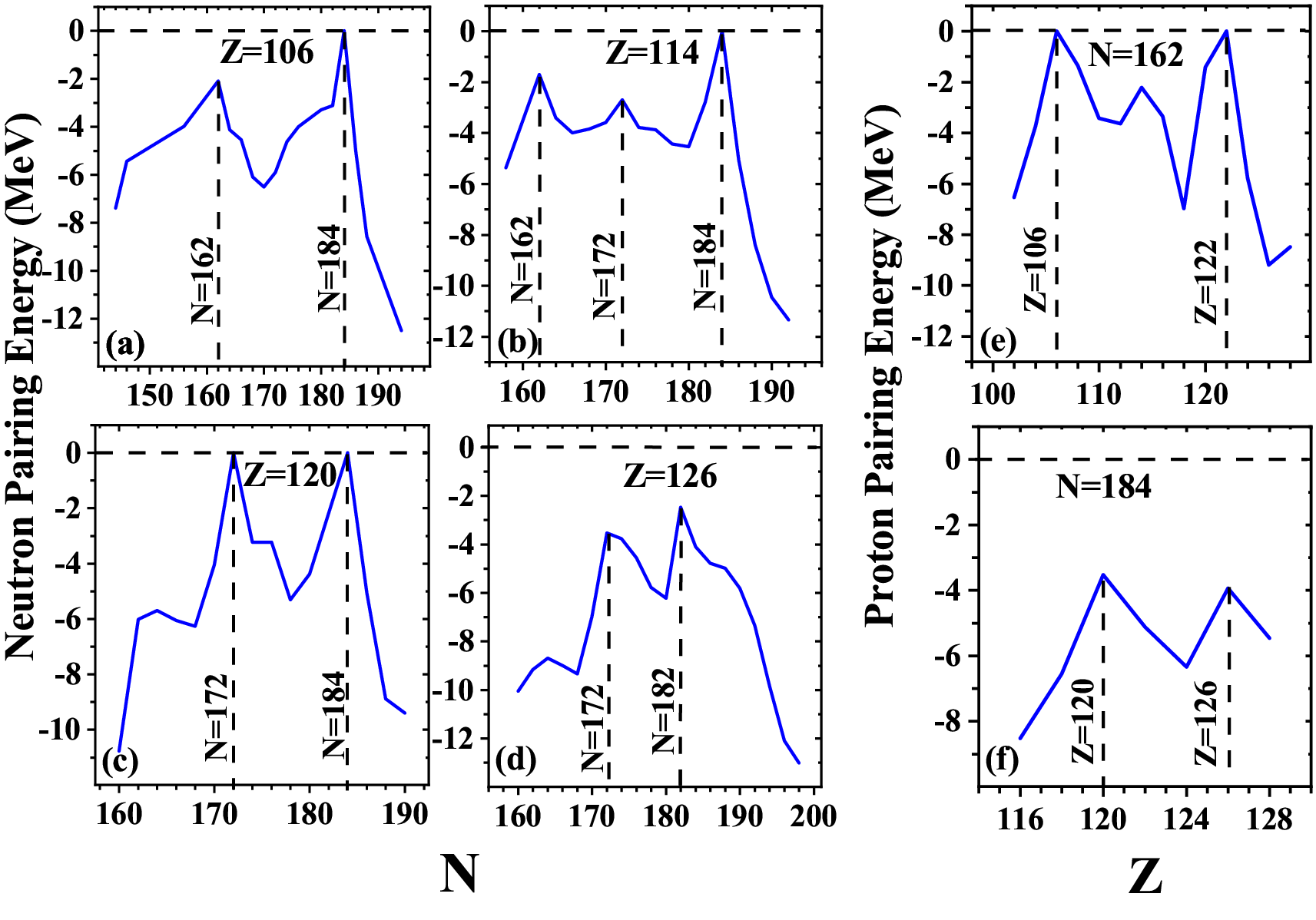}
\caption{(Colour online) Pairing energies (MeV) for Z$=$106, 114, 120, 126 and for N$=$162, 184 from RMF approach using DD-ME2 \cite{Lalazissis05} force parameter. }\label{pairing}
\end{figure*}

To provide further support, the contribution of pairing energy, which plays a crucial role in the stability of nuclei, is depicted in Fig. \ref{pairing}. The pairing energy curve develops a peak when the corresponding nucleus exhibits greater stability compared to its neighbors. In some cases, the curve vanishes, indicating the presence of a shell closure. From Fig. \ref{pairing}, lower pairing energy values suggest the magic behavior of N$=$162, 172, 184, along with Z$=$ 106, 114, 120, 126, etc. Notably, zero neutron pairing energy for $^{290}106$, $^{298}114$, $^{292}120$, and $^{304}120$ points towards strong magicity and provides substantial evidence for the existence of the next doubly magic nuclei beyond $^{208}$Pb.\par

\begin{figure*}[!htbp]
\centering
\includegraphics[width=0.8\textwidth]{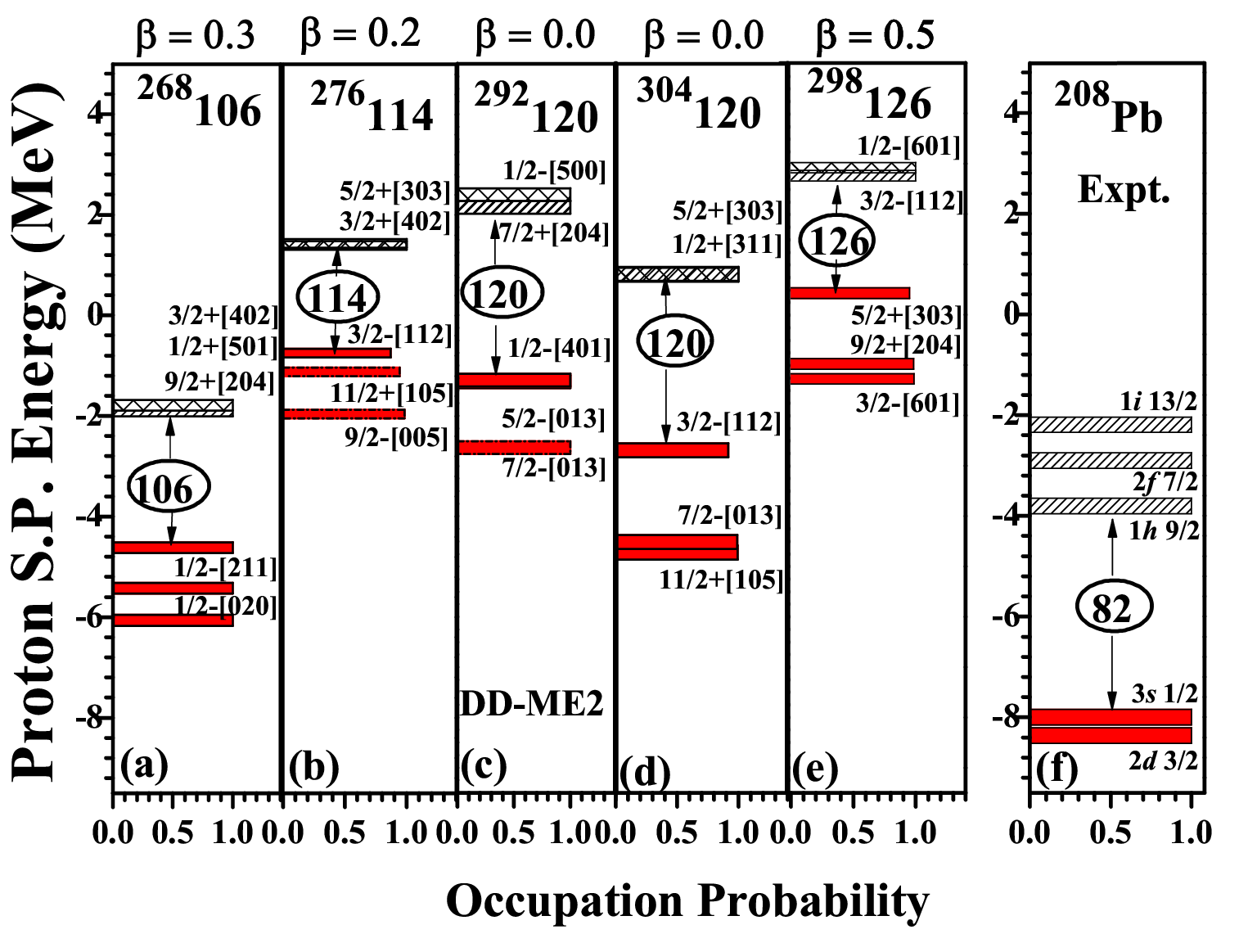}
\caption{(Colour online) Proton single particle energy for nuclei with Z$=$106, 114, 120, and 126 obtained from RMF approach using DD-ME2 \cite{Lalazissis05} parameter.}\label{spe}
\end{figure*}

The nucleus $^{208}$Pb is a well-known and very strong magic candidate, exhibiting a spherical nature with a closed shell for both neutrons and protons. However, in the superheavy region, the presence of deformation makes the spherical magic nucleus less likely. Nonetheless, some nuclei with neutron numbers N$=$172 and several nuclei with N$=$184 (see Fig. \ref{pes}) are found to be spherical. In the search for double magicity, it is crucial to examine the shell gap in the proton distribution. In magic nuclei, the shell gap refers to the energy difference between the highest occupied nuclear shell and the next unoccupied shell, according to the nuclear shell model. This gap is substantially larger in nuclei with specific numbers of protons or neutrons, known as magic numbers (e.g., 2, 8, 20, 28, 50, 82, 126). These large gaps confer enhanced stability, as a significant amount of energy is required to add or remove nucleons, making magic nuclei particularly resistant to nuclear reactions and decay. To explore this concept, we have presented the proton Nilsson single-particle levels (accounting for the substantial deformation in superheavy nuclei) and their occupancy near the Fermi energy for selected nuclei with $Z = 106, 114, 120$, and $126$ in Fig. \ref{spe}. The respective ground-state deformations are indicated at the top of each panel. Filled states are highlighted in red, corresponding to their occupancy, while empty states are represented by shaded bars. For reference, we have also included the experimental proton single-particle levels of $^{208}\text{Pb}$ \cite{nndc}. A clear shell gap of around 4 MeV is evident in panel (f) of Fig. \ref{spe}, accounting for the magic behavior of proton number $Z = 82$. Similarly, a significant energy gap ($\sim$ 3 MeV) between the levels 1/2$^{-}$[401] (3/2$^{-}$[112]) and 7/2$^{+}$[204] (1/2$^{+}$[311]) results in magicity at $Z = 120$ in nuclei like $^{292}120$ and $^{304}120$, establishing their strong magic nature with a spherical shape. It is noteworthy that all the filled states are fully occupied (Occupation probability $=$ 1), and the states above the gap are fully empty, providing further evidence for the shell closure at Z$=$120. A similar, though less pronounced, gap can be observed for Z$=$106, 114, and 126, where the filled states are almost fully occupied, and others are empty, even with their non-zero deformation. This explains why the proton pairing energy peaks for Z$=$106, Z$=$114, and Z$=$126 in Figs. \ref{pairing} (e) and (f). \par

\begin{figure}[h!]
    \centering
    \includegraphics[width=0.5\textwidth, height=0.3\textwidth]{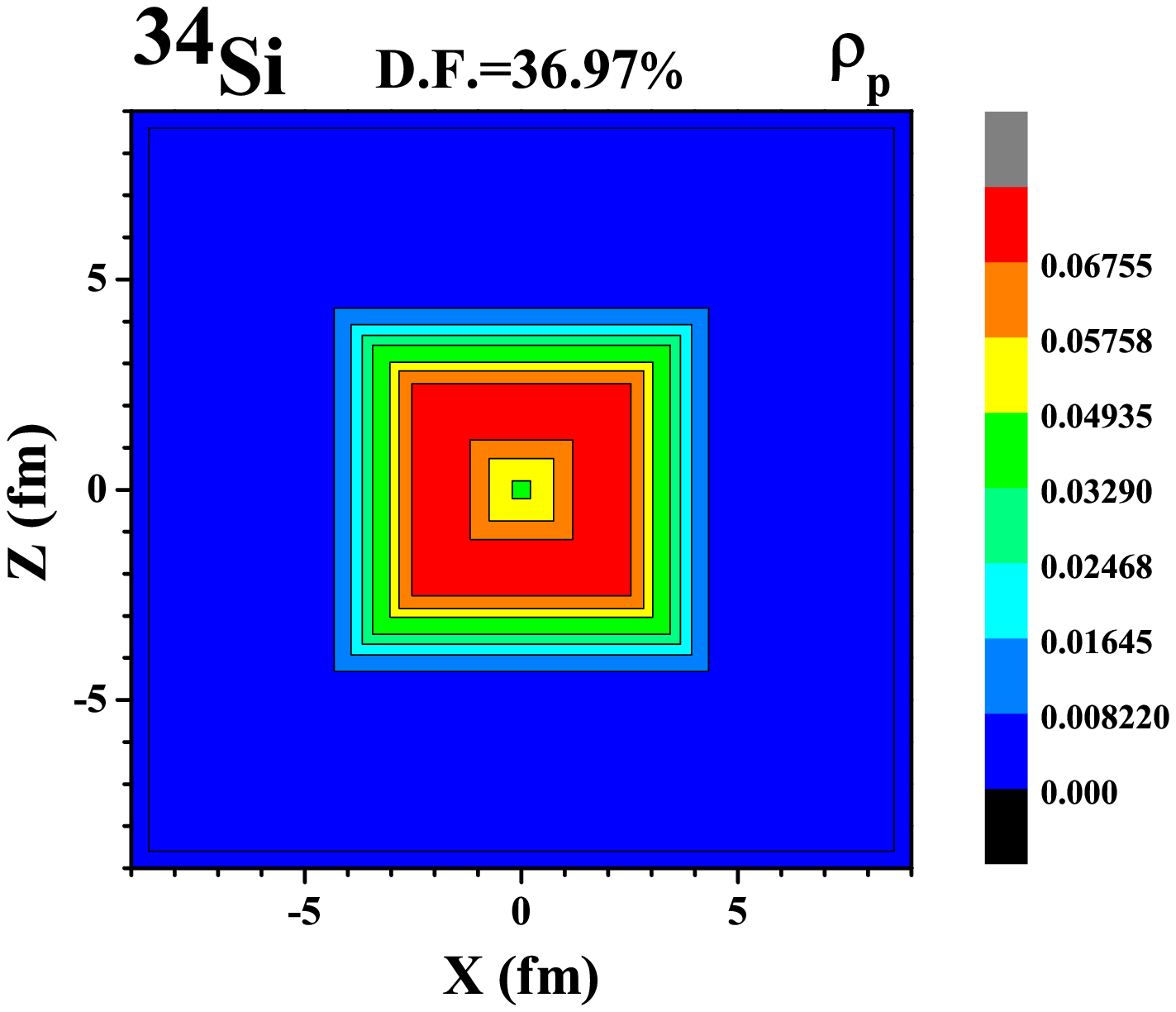}  
\vspace{0.2cm}  
    \includegraphics[width=0.8\textwidth]{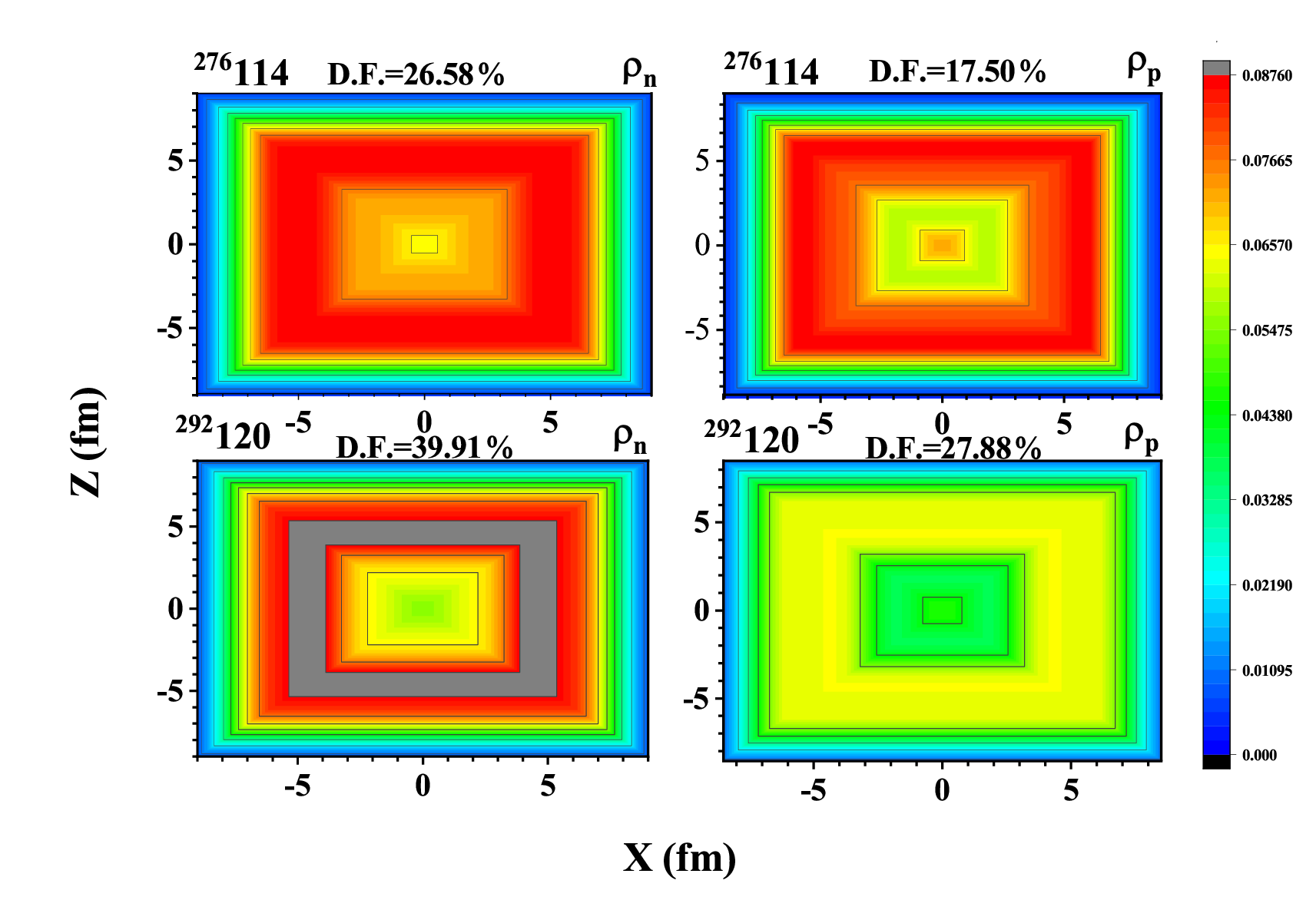}  
    \label{fig:figure2}
    \caption{(Colour online) Radial density distribution of proton and neutron densities for $^{276}114$ and $^{292}120$. Respective D.F. is also mentioned at the top of the panel.}
    \label{PND}
\end{figure}

Another significant and interesting feature of magic nuclei is the depletion in their nucleonic central density, a phenomenon observed in the case of $^{34}$Si, one of the magic isotones of N$=$20. This depletion is speculated to be a general feature in several magic nuclei \cite{saxena}. In the superheavy region, the depletion in proton density is primarily attributed to the Coulomb repulsion of protons \cite{saxenaijmpe2018,saxenaijmpe2019,saxena}. This depletion is found to depend on neutron and proton numbers. The depletion in central density, often referred to as the 'bubble structure' in SHN, has been studied in a few isotopes of Z$=$120 \cite{li2016,schuetrumpf2017}. It is associated with the large repulsive Coulomb field and strong centrifugal potential due to the presence of high-$l$ states, which drive the distribution of protons toward the surface of the nucleus. \par


\begin{figure}[h]
\centering
\includegraphics[width=0.8\textwidth]{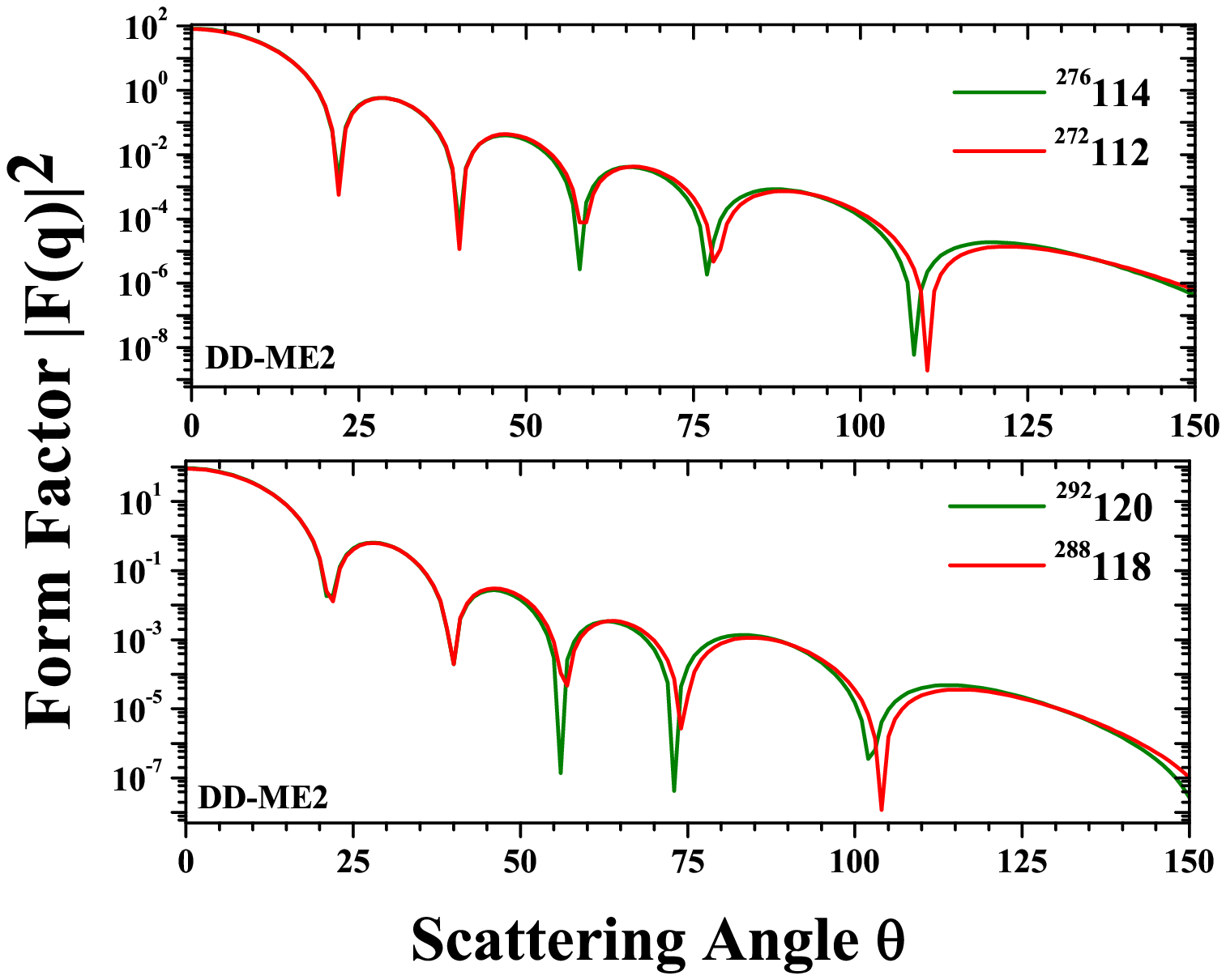}
\caption{(Colour online) Nuclear charge form factors vs. scattering angle for $^{276}114$ and $^{292}120$ and their neighbour nuclei.}\label{ff}
\end{figure}

In Fig. \ref{PND}, we have presented the radial density distribution of neutron and proton density ($\rho_{n}$ and $\rho_{p}$) for Z=114 with N=162, along with Z=120 with N=172. To examine the variation in central depletion, quantified in terms of the depletion fraction (D.F.=$(\rho_{max}-\rho_{c})/\rho_{max}$, where $\rho_{max}$ and $\rho_{c}$ are the maximum and central densities, respectively), the respective D.F. is also indicated at the top of the panel. A substantial formation of bubbles for neutrons and protons is observed for both nuclei. Hence, among superheavy magic nuclei, $^{276}114$ and $^{292}120$ are reported here as potential bubble candidates. As a reference, we have shown the proton density distribution of \(^{34}\text{Si}\), which currently provides the only experimental evidence of density depletion. This depletion is attributed to the vacant single-particle \( s \)-state. In contrast, the bubble structures observed in the superheavy nuclei examined in this study are primarily driven by central Coulomb repulsion. Although the central depletion in \(^{34}\text{Si}\) is quite pronounced, its radial density distribution is not as extensive as that observed in superheavy nuclei. Nevertheless, this comparison provides valuable insights into bubble structures across different nuclear mass regions. The significant value of the depletion factor (D.F.) for \(^{292}120\) can be attributed to its magic nature, which is found to be substantial for \( N = 172 \) with a proton single-particle gap similar to that observed in \(^{292}120\) (see Fig. \ref{spe}). In addition, the presence of a higher $l$ state viz. $l\ge$4 for $^{292}120$ and $l\ge$5 for $^{276}114$ near the Fermi level supports the formation of central depletion. Hence, the bubble formation in SHN is somewhat sensitive to its magic character as it is found in other magic nuclei of the periodic chart in our earlier investigation \cite{saxena}.\par

The nuclear charge form factor, a useful physical observable of central depletion, is a measurable quantity through elastic electron-nucleus scattering experiments~\cite{hofstadter,forest,donnelly}. The form factors for $^{276}114$ and $^{292}120$, as well as their neighboring nuclei, namely $^{272}112$ and $^{288}118$, are depicted in Fig. \ref{ff}. The systematic study presented above does not indicate any magic character in the nuclei with Z=112 and 118, whereas nuclei with Z=114 and 120 exhibit a strong magic character. This distinction is evident in the variation of our computed charge density form factor, as shown in Fig. \ref{ff}, where the third, fourth, and fifth peaks of the form factor are distinctly separated for Z=114 and 120, indicating the presence of bubble and non-bubble nuclei. This information is closely related to the magic and non-magic characters of these nuclei. A more detailed study, considering the effects of pairing and deformation, may provide further insights into the complex phenomenon of central depletion and its relation to magicity in SHN.\par

\subsection{$\alpha$-decay chains}
As mentioned in the previous subsection, there are a few SHN that exhibit magic characters, indicating more stable configurations compared to their neighboring nuclei. Therefore, these nuclei are the most probable candidates expected to be synthesized in the laboratory in the future.

\begin{figure}[h]
\centering
\includegraphics[width=0.65\textwidth]{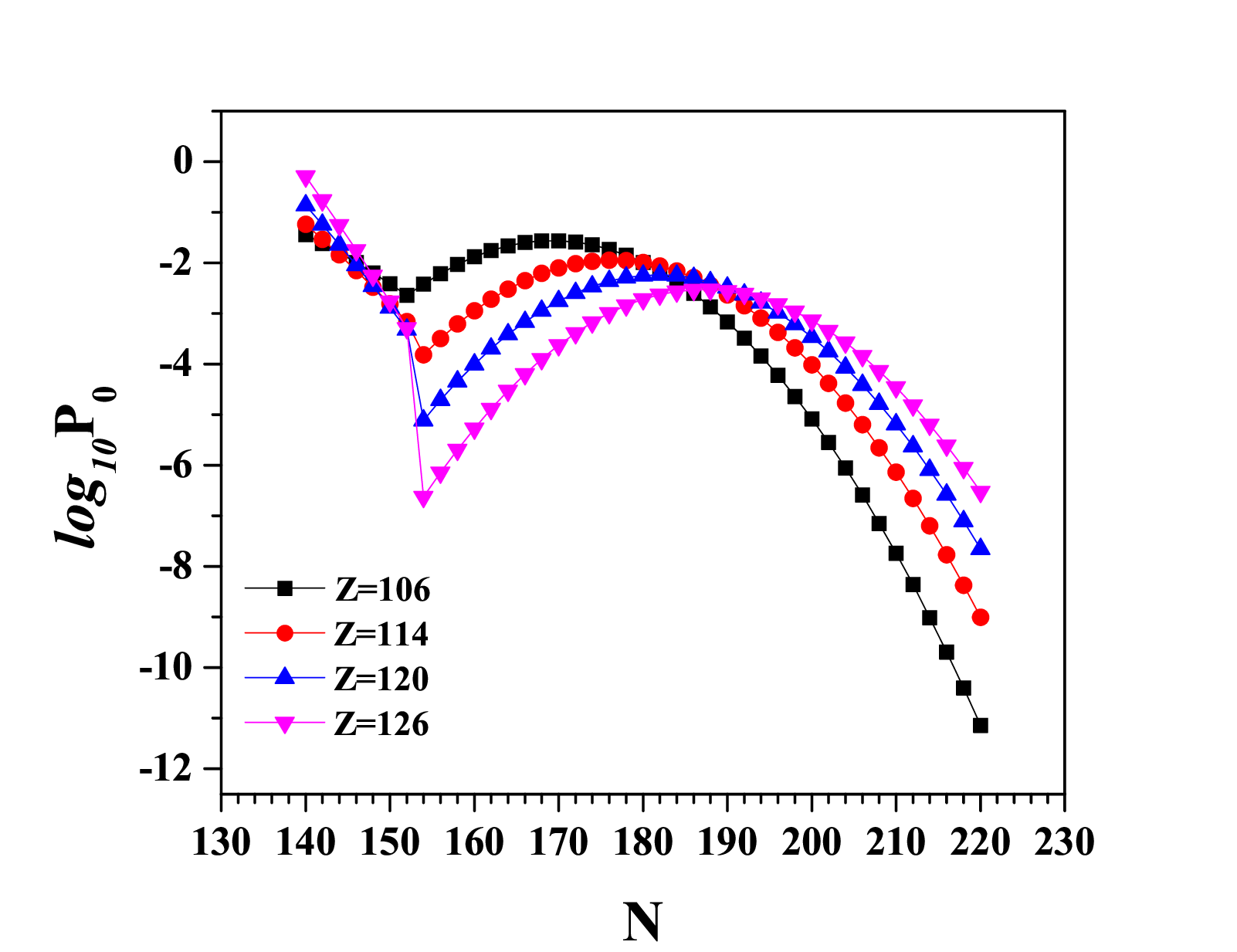}
\caption{(Colour online) $\alpha$-particle preformation probability P$_0$ \cite{guo2015} for Z$=$106, 114, 120, and 126 isotopes.}\label{p0}
\end{figure}

As seen in Fig. \ref{qalpha}, a certain change in $\alpha$-decay energy (Q${\alpha}$) from their systematics indicates the stability of SHN nuclei. Therefore, the study of $\alpha$-decay half-lives is crucial for these magic nuclei, considering that $\alpha$-decay half-lives primarily depend on Q${\alpha}$ values. In this subsection, we analyze the decay properties and half-lives of the considered SHN. $\alpha$-decay is a precise tool for investigating detailed nuclear structures, such as ground-state energy, effective nuclear interaction, shell effects, nuclear spin, parity, etc. \cite{horiuchi1991,lovas1998,hodgeson2003}. In this regard, the $\alpha$-particle preformation probability P$_0$ is of enormous importance as it interprets the preformation as the pre-scission part of barrier penetration and directly relates to information on nuclear structure \cite{dussel1986,gambhir1983,kaneko2003}. We use the recently reported empirical formula by Guo \textit{et al.} \cite{guo2015} to evaluate P$_0$ for Z$=$106, 114, 120, and 126 isotopic chains, as shown in Fig. \ref{p0}. The large values of P$_0$ around N$\sim$184, followed by a drop towards both sides of the peak, are a clear signature of a higher probability of $\alpha$-decay for the region N$\sim$184 in the considered isotopic chains. In particular, the maximum probability of $\alpha$-decay is observed at N$=162$, N$=172$, N$=184$, and N$=190$ for Z$=$106, 114, 120, and 126, respectively.\par

\begin{figure*}[h]
\centering
\includegraphics[width=0.8\textwidth]{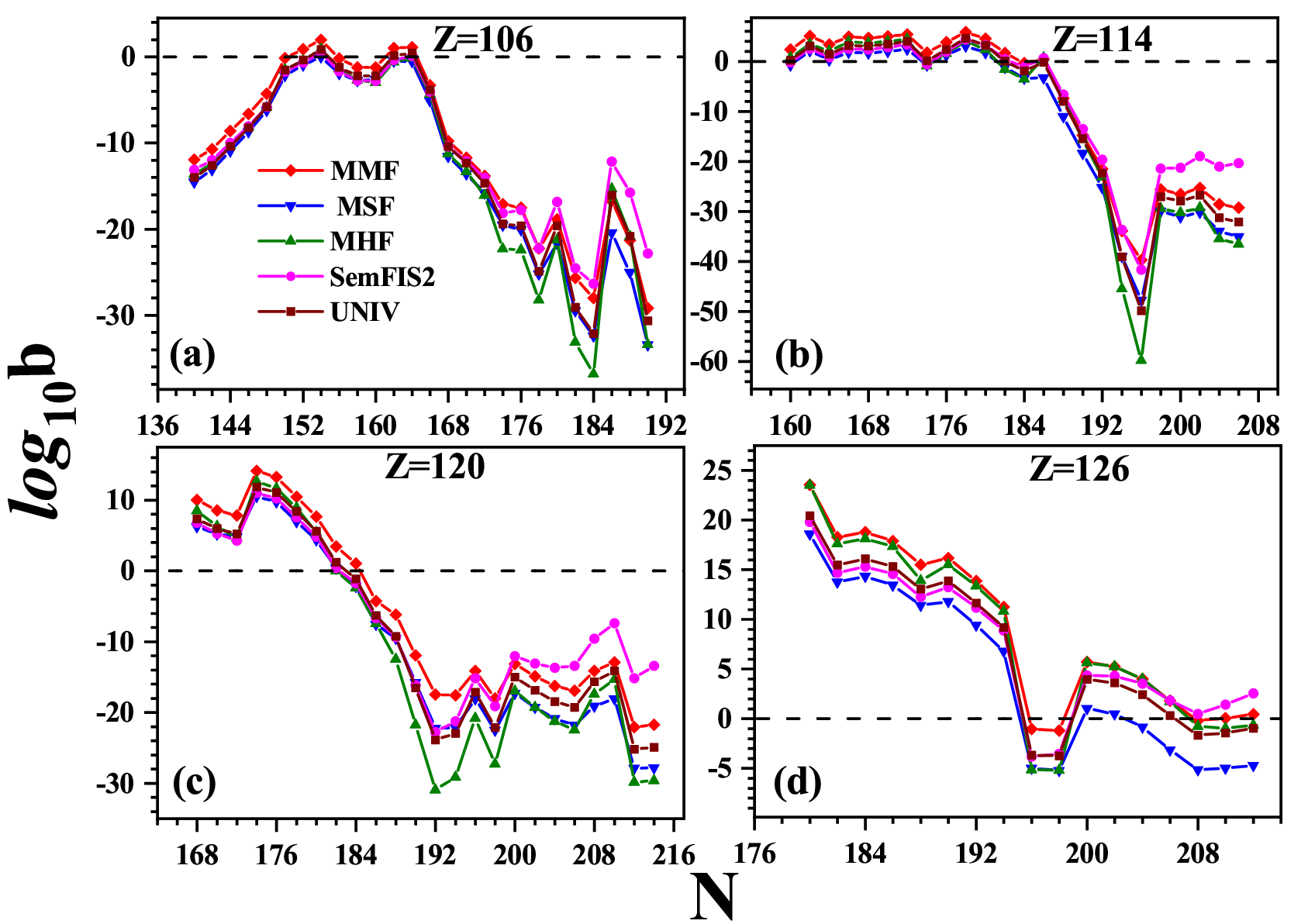}
\caption{(Colour online) Branching ratio ($log_{10}$b) for Z$=$106, 114, 120, and 126 isotopes.}\label{alpha-SF}
\end{figure*}

To confirm the likelihood of $\alpha$-decay, we examine the competition between the dominant decay modes in this region, namely $\alpha$-decay and SF. We quantify this competition by defining the branching ratio as \cite{pathak}:

\begin{equation}
b = \frac{T^{SF}_{1/2}}{T^{\alpha}_{1/2}}
\end{equation}

The branching ratio 'b' is defined as the ratio of $\alpha$-decay half-lives ($T^{\alpha}_{1/2}$) to spontaneous fission half-lives ($T^{SF}_{1/2}$). For the present work, $T^{\alpha}_{1/2}$ is calculated using five empirical formulas mentioned in subsection \ref{decay}. These include the recent MMF \cite{singhnpa2020}, MHF \cite{saxena2020prc}, and MSF \cite{saxena2020prc} formulas, along with the widely used SemFIS2 \cite{poenaru2007} and UNIV \cite{poenaru2011} formulas. The values of $\alpha$-decay half-lives using these formulas are listed in Table \ref{hl} of the appendix for Z$=$106, 114, 120, and 126 isotopes. The spontaneous fission half-life $T^{SF}_{1/2}$ is calculated using MBF-2020 formula \cite{saxena2020prc}. The logarithm value of the branching ratio 'b' is plotted in Fig. \ref{alpha-SF} for the considered chains of isotopes, revealing a clear competition between $\alpha$-decay and SF, which delimits the upper limit of $\alpha$-transition for the concerned Z.\par

Nuclei for which the $log_{10}$b value is greater than zero would resist SF, making them likely to decay by emitting an $\alpha$-particle and thus suitable for experimental investigation. From the figure, N$\sim$164 and N$\sim$192 represent upper limits for $\alpha$-transition in Z$=$106 and 126 isotopes, respectively. Similarly, N$\sim$182 serves as the upper limit for $\alpha$-transition in both Z$=$114 and 120 isotopes, which closely aligns with the expectations from $\alpha$-particle preformation probability P$_0$ (Fig. \ref{p0}). Among the four considered proton magic Z, we propose eight $\alpha$-decay chains starting from Z$=$126, and these chains automatically include nuclei with other considered magic numbers Z$=$120, 114, and 106. Probable decay modes are also compared with experimental estimations \cite{nndc}, showing excellent agreement. We observe long $\alpha$-decay chains for the region around N$\sim$184 as mentioned in Table \ref{DMT} of the appendix, confirming the stability in that vicinity.

\section{Conclusions}
We have employed the state-of-the-art relativistic mean-field (RMF) approach for an extensive and systematic study of nuclei with proton magic numbers 106, 114, 120, 126, and neutron magic numbers 162, 172, and 184 in the superheavy region. Results of ground-state properties such as binding energy, deformation, shape, radii, and charge density show excellent agreement with available experimental data. We predict the central depletion in charge density as a general phenomenon in SHN which is due to strong Coulomb repulsion as well as large centrifugal force originating from occupied higher $l$-orbits. Decay properties of the isotopic chain of Z$=$106, 114, 120, and 126 are presented to identify possible decay modes by comparing the $\alpha$-decay half-lives and the corresponding SF half-lives. Moreover, long $\alpha$-decay chains are anticipated near the island of stability N$\sim$184. This study may provide useful inputs for future experiments and the search for new elements in the superheavy region.
\section*{Acknowledgement}
AJ expresses deep gratitude to Prof. S. K. Jain for his unwavering guidance. The authors also thank P. K. Sharma (GPC Rajasamand, Rajasthan, India), S. Agarwal (J.D.B. Girls College, Kota, Rajasthan, India), and S. Swami (Shekhawati University, Rajasthan, India) for their valuable contributions.

\onecolumn

\section{Appendix}
\begin{table*}[!htbp]
\caption{The predicted proton and neutron magic numbers based on several theories and models.\label{magic}}
\centering
\resizebox{0.7\textwidth}{!}{%
{\begin{tabular}{|c|l|}
\hline
\multicolumn{2}{|c|}{Proton magic nuclei}\\
\hline
\multicolumn{1}{|c|}{Z}&
\multicolumn{1}{c|}{Corresponding neutron number N}\\

 \hline
 106&184, 194, 224 \cite{Ismail2016}\\
108&162\cite{Zhang2012,Ismail2016}, 168, 184, 190, 220\cite{Ismail2016}\\
110&162,184\cite{Li2017}\\
112&162,184\cite{Li2017}\\
 114&160,\cite{Seyyedi2a0}162,\cite{Li2017}164,\cite{Seyyedi2a0} 170,\cite{Seyyedi2a0} 178,\cite{Ismail2015} \\
 &184, \cite{Li2017,Manju2020,Rutz1997,Seyyedi2a0,Zhang2012}, 196,\cite{Manju2020} 228,236,274\cite{Ismail2015}\\
116&164,174,176\cite{Seyyedi2a0}\\
118&176,\cite{Seyyedi2a0} 228\cite{saxena1}\\
120&  164, \cite{Zhang2005} 228,\cite{Zhang2005,saxena1} \cite{Zhang2005}
172,\cite{Rutz1997,Zhang2005,Seyyedi2a0,Patra2014,Ismail2017,Beckmann2000,Patra2021,Patra2012} 182,\cite{Patra2012} 184,\cite{Manju2020,Patra2014,Ismail2017,Patra2021,Beckmann2000,Bezbakh15,Patra2012,Zhang2005}\\
&198,\cite{Beckmann2000,Zhang2005} 200, 206, \cite{Ismail2016}208,\cite{Patra2012} 228,\cite{Zhang2005} 250\cite{Ismail2015} 252,\cite{Zhang2005}
256,\cite{Ismail2016} 258\cite{Patra2012,Zhang2005,Ismail2015} 274,\cite{Zhang2005}\\
122& 168, 172, 174, 178, 184,\cite{Siddiqui2020} 206, 216,\cite{Ismail2016} 228,\cite{saxena1,Ismail2016} 238, 246\cite{Ismail2016}\\
124 &168,\cite{Siddiqui21} 174,\cite{Siddiqui21,Siddiqui2020} 178,\cite{Siddiqui21,Siddiqui2020} 184,\cite{Kruppa00,Cwiok96,Cwiok05,Siddiqui2020}
186\cite{Manju2020}\\
126 &174,178\cite{Siddiqui21} 184,\cite{Manju2020,Rutz1997,Kruppa00,Cwiok96,Cwiok05} 196,\cite{Manju2020}\\
128&168, 174, 178,182,200\cite{Siddiqui2020}\\
132& 172, 184, 228, 238, 258. \cite{Zhang2005}\\
138& 164, 172, 184, 198, 228, 238, 252, 258, 274 \cite{Zhang2005}\\
\hline
\hline
\multicolumn{2}{|c|}{Neutron magic nuclei}\\
\hline
\multicolumn{1}{|c|}{N}&
\multicolumn{1}{c|}{Corresponding proton number Z}\\
\hline
138&106,108\cite{Zhang2005}\\
152&102,\cite{singhnpa2020}104,106,108\cite{Li2017}\\
162&104,\cite{Li2017}106,\cite{Li2017,Ghodsi2020}108,\cite{Li2017,singhnpa2020,Ghodsi2020,Zhang2012}110,112,\cite{Ghodsi2020}
114,\cite{Ghodsi2020,Li2017} 178,182,120,\cite{Ghodsi2020}\\
172  &120,\cite{Siddiqui2020,Zhang2005}132,138 \cite{Zhang2005}\\
178  &116,118,120\cite{Ghodsi2020}\\
184  &104,\cite{singhnpa2020} 114,\cite{Zhang2012,Ghodsi2020,Li2017,Manju2020} 120,\cite{Ghodsi2020,Zhang2005,Siddiqui2020}
132,138 \cite{Zhang2005}124,126,\cite{Manju2020}\\
186&124\cite{Manju2020}    \\
196&114, 122, 126\cite{Manju2020} \\
198  &120,132,138 \cite{Zhang2005}\\
216 &122,\cite{Ismail2016}\\
228    &120,132,138 \cite{Zhang2005}\\
238     &120,132,138 \cite{Zhang2005}\\
258      &120,132,138 \cite{Zhang2005}\\
\hline
\end{tabular}}}
\end{table*}
\onecolumn
\begin{center}
\small
\setlength{\tabcolsep}{3pt}
\begin{longtable}{|c|c|c|c|c|c|c|c|c|}
\caption{$\alpha$-decay chains for unknown SHN starting from Z$=$126 are listed. SHN with lower $\alpha$-fission half-life would decay with the emission of $\alpha$-particle while for other nuclei SF would be dominant. Experimental (Expt.) decay mode are taken from Ref. \cite{nndc}.\label{DMT}}\\
\hline
 \multicolumn{1}{|c}{Nucleus}&
 \multicolumn{1}{|c}{$log_{10}T^{SF}_{1/2}(s)$}&
 \multicolumn{5}{|c}{{$log_{10}T^{\alpha}_{1/2}(s)$}}&
 \multicolumn{2}{|c|}{Decay Mode} \\
 \cline{2-9}
 \multicolumn{1}{|c}{}&
 \multicolumn{1}{|c}{MBF}&
  \multicolumn{1}{|c}{MMF}&
   \multicolumn{1}{|c}{MSF}&
  \multicolumn{1}{|c}{MHF}&
   \multicolumn{1}{|c}{SemFIS2}&
 \multicolumn{1}{|c}{UNIV}&
 \multicolumn{1}{|c}{Predicted}&
   \multicolumn{1}{|c|}{Expt.}\\
\hline
\endfirsthead
\multicolumn{9}{c}%
{\tablename\ \thetable\ -- \textit{Continued from previous page}} \\
\hline
 \multicolumn{1}{|c}{Nucleus}&
 \multicolumn{1}{|c}{$log_{10}T^{SF}_{1/2}(s)$}&
 \multicolumn{5}{|c}{{$log_{10}T^{\alpha}_{1/2}(s)$}}&
 \multicolumn{2}{|c|}{Decay Mode} \\
 \cline{2-9}
 \multicolumn{1}{|c}{}&
 \multicolumn{1}{|c}{MBF}&
  \multicolumn{1}{|c}{MMF}&
   \multicolumn{1}{|c}{MSF}&
  \multicolumn{1}{|c}{MHF}&
   \multicolumn{1}{|c}{SemFIS2}&
 \multicolumn{1}{|c}{UNIV}&
 \multicolumn{1}{|c}{Predicted}&
   \multicolumn{1}{|c|}{Expt.}\\
\hline
\endhead
\hline \multicolumn{9}{r}{\textit{Continued on next page}} \\
\endfoot
\hline
\endlastfoot
$^{	306	}$126&11.22  & -12.33 & -7.39 & -12.31  & -8.60  & -9.23 &  $ \alpha $  &              \\
$^{	302	}$124&9.09   & -9.16  & -4.90 & -8.21   & -5.68  & -6.43 &  $ \alpha $  &              \\
$^{	298	}$122&6.45   & -8.60  & -4.54 & -7.63   & -5.34  & -6.06 &  $ \alpha $  &              \\
$^{	294	}$120&6.69   & -6.45  & -2.87 & -4.96   & -3.37  & -4.14 &  $ \alpha $  &              \\
$^{	290	}$118&4.50   & -3.93  & -0.85 & -1.81   & -0.94  & -1.77 &  $ \alpha $  &              \\
$^{	286	}$116&2.97   & -6.92  & -3.49 & -6.01   & -4.36  & -5.00 &  $ \alpha $  &              \\
$^{	282	}$114&-0.95  & -5.70  & -2.61 & -4.68   & -3.38  & -4.01 &  $ \alpha $  &              \\
$^{	278	}$112&-1.79  & -4.87  & -2.05 & -3.88   & -2.81  & -3.41 &  $ \alpha $  &              \\
$^{	274	}$110&4.96   & -2.51  & -0.22 & -1.20   & -0.66  & -1.26 &  $ \alpha $  &              \\
$^{	270	}$108&4.52   & -4.28  & -1.82 & -3.70   & -2.81  & -3.29 &  $ \alpha $  &  $ \alpha $  \\
$^{	266	}$106&0.62   & 1.82   & 3.24  & 3.60    & 3.40   & 2.83  &  SF          &  SF          \\
\hline
$^{	308	}$126&8.38   & -9.89  & -5.38 & -9.25   & -6.31  & -7.08 & $ \alpha $     &            \\
$^{	304	}$124&3.83   & -3.86  & -0.32 & -0.94   & -0.09  & -1.13 & $ \alpha $     &            \\
$^{	300	}$122&4.88   & -10.88 & -6.35 & -10.86  & -7.70  & -8.31 & $ \alpha $     &            \\
$^{	296	}$120&7.61   & -5.67  & -2.16 & -4.10   & -2.67  & -3.47 & $ \alpha $     &            \\
$^{	292	}$118&6.14   & -4.31  & -1.12 & -2.51   & -1.49  & -2.29 & $ \alpha $     &            \\
$^{	288	}$116&3.40   & -4.85  & -1.71 & -3.53   & -2.36  & -3.06 & $ \alpha $     &            \\
$^{	284	}$114&0.13   & -4.91  & -1.89 & -3.90   & -2.73  & -3.35 & $ \alpha $     &  SF        \\
$^{	280	}$112&-5.10  & -3.99  & -1.24 & -2.99   & -2.06  & -2.65 &  SF            &            \\
$^{	276	}$110&2.08   & -2.44  & -0.06 & -1.32   & -0.75  & -1.32 & $ \alpha $     &            \\
$^{	272	}$108&3.58   & -2.69  & -0.41 & -1.99   & -1.36  & -1.83 & $ \alpha $     &            \\
$^{	268	}$106&2.98   & 1.95   & 3.46  & 3.57    & 3.33   & 2.82  & $ \alpha $/SF  &            \\
\hline
$^{	310	}$126&9.52   &-9.26  & -4.82  & -8.60  & -5.81   & -6.58 & $ \alpha $     &                \\
$^{	306	}$124&2.01   &-1.89  & 1.52   & 1.85   & 2.04    & 0.93  & $ \alpha $     &                \\
$^{	302	}$122&4.29   &-9.73  & -5.37  & -9.59  & -6.72   & -7.35 & $ \alpha $     &                \\
$^{	298	}$120&5.25   &-5.23  & -1.72  & -3.70  & -2.36   & -3.14 & $ \alpha $     &                \\
$^{	294	}$118&6.15   &-4.04  & -0.82  & -2.35  & -1.37   & -2.13 & $ \alpha $     &                \\
$^{	290	}$116&4.41   &-1.14  & 1.60   & 1.38   & 1.52    & 0.74  & $ \alpha $     &  $ \alpha$     \\
$^{	286	}$114&1.16   &-4.28  & -1.27  & -3.32  & -2.26   & -2.84 & $ \alpha $     & SF /$ \alpha$  \\
$^{	282	}$112&-4.26  &-3.03  & -0.33  & -1.98  & -1.26   & -1.80 &  SF            &                \\
$^{	278	}$110&-4.39  &-1.68  & 0.69   & -0.57  & -0.16   & -0.66 &  SF            &                \\
$^{	274	}$108&0.45   &-1.07  & 1.07   & -0.18  & 0.13    & -0.30 & $ \alpha $/SF  &                \\
$^{	270	}$106&2.74   &1.59   & 3.26   & 2.88   & 2.66    & 2.28  & $ \alpha $/SF   &                \\
\hline
$^{	312	}$126&8.90  &-9.00 &-4.55 &-8.47 &-5.70 &-6.44  &$\alpha$    &                \\
$^{	308	}$124&-0.27 &-1.66 &1.79  &2.03  &2.11  &1.07   & SF         &                \\
$^{	304	}$122&2.85  &-8.79 &-4.53 &-8.58 &-5.96 &-6.58  & $ \alpha $ &                \\
$^{	300	}$120&4.72  &-2.99 &0.30  &-0.73 &-0.11 &-0.90  & $ \alpha $ &                \\
$^{	296	}$118&5.65  &-3.97 &-0.68 &-2.48 &-1.51 &-2.18  & $ \alpha $ &                \\
$^{	292	}$116&4.81  &-1.74 &1.14  &0.30  &0.60  &-0.06  & $ \alpha $ & $ \alpha $     \\
$^{	288	}$114&2.42  &0.65  &3.16  &3.31  &2.95  &2.31   & $ \alpha $ & $ \alpha $     \\
$^{	284	}$112&-2.49 &-2.32 &0.38  &-1.29 &-0.74 &-1.19  & SF         & SF             \\
$^{	280	}$110&-5.25 &-0.70 &1.66  &0.49  &0.64  &0.25   & SF         &                \\
$^{	276	}$108&-4.96 &0.77  &2.82  &2.03  &1.87  &1.55   &  SF        &                \\
$^{	272	}$106&-1.75 &1.53  &3.34  &2.56  &2.29  &2.07   &  SF        &                \\
\hline
$^{	314	}$126&9.47  &-6.04  &-1.96 &-4.45  &-2.79 &-3.59  & $ \alpha $  &              \\
$^{	310	}$124&2.19  &-5.55  &-1.64 &-4.02  &-2.54 &-3.28  & $ \alpha $  &              \\
$^{	306	}$122&0.39  &-10.63 &-5.94 &-11.32 &-7.93 &-8.40  & $ \alpha $  &              \\
$^{	302	}$120&3.41  &-0.08  &3.03  &3.42   &2.99  &2.21   & $\alpha $/SF&              \\
$^{	298	}$118&5.18  &-3.62  &-0.28 &-2.23  &-1.38 &-1.93  & $ \alpha $  &              \\
$^{	294	}$116&5.75  &-1.65  &1.33  &0.20   &0.43  &-0.08  &  $ \alpha $ &              \\
$^{	290	}$114&4.16  &0.28   &2.93  &2.55   &2.22  &1.76   &  $ \alpha $ &              \\
$^{	286	}$112&-0.60 &2.61   &4.93  &5.47   &4.51  &4.13   &  SF         &              \\
$^{	282	}$110&-4.17 &0.15   &2.54  &1.40   &1.26  &1.03   &  SF         &              \\
$^{	278	}$108&-7.12 &1.71   &3.79  &3.07   &2.58  &2.45   & SF          &              \\
$^{	274	}$106&-6.37 &3.38   &5.18  &4.86   &4.04  &4.02   & SF          &              \\
\hline
$^{	316	}$126&8.66 &-7.53 &  -3.16 &-6.85  &-4.59 &-5.21 & $ \alpha $&         \\
$^{	312	}$124&1.11 &3.32  &  6.77  &9.82   &7.77  &6.80  & SF&         \\
$^{	308	}$122&-4.89&-10.46&	 -5.71 &-11.39 &-7.98 &-8.36 & $ \alpha $&         \\
$^{	304	}$120&1.20 &0.15  &  3.34  &3.57   &2.94  &2.35  & SF        &         \\
$^{	300	}$118&3.98 &-2.16 &  1.11  &-0.37  &-0.08 &-0.50 & $ \alpha $&         \\
$^{	296	}$116&5.74 &-1.02 &  2.00  &0.87   &0.80  &0.47  & $ \alpha $&         \\
$^{	292	}$114&5.71 &-0.20 &  2.62  &1.60   &1.33  &1.09  & $ \alpha $&         \\
$^{	288	}$112&-0.16&2.62  &  5.08  &5.26   &4.15  &4.00  & SF        &         \\
$^{	284	}$110&-2.83&5.69  &  7.87  &9.35   &7.38  &7.36  & SF        &         \\
$^{	280	}$108&-6.01&2.85  &  4.99  &4.44   &3.49  &3.61  & SF        &         \\
$^{	276	}$106&-7.43&4.27  &  6.17  &5.88   &4.65  &4.91  & SF        &         \\
\hline
$^{	318	}$126&6.49  &-7.38 &-2.95 &-6.90  &-4.70  &-5.18   & $ \alpha $ &        \\
$^{	314	}$124&2.98  &-1.78 &1.91  &1.22   &1.10   &0.55    & $ \alpha $ &        \\
$^{	310	}$122&0.11  &-13.82&-8.21 &-16.12 &-11.18 &-11.38  & $ \alpha $ &        \\
$^{	306	}$120&-4.07 &0.17  &3.47  &3.37   &2.59   &2.23    & SF         &        \\
$^{	302	}$118&1.96  &-10.73&-5.87 &-12.43 &-8.89  &-9.01   & $\alpha$   &        \\
$^{	298	}$116&5.01  &0.53  &3.54  &2.92   &2.18   &2.06    & $\alpha$   &        \\
$^{	294	}$114&5.69  &1.10  &3.93  &3.27   &2.41   &2.42    & $\alpha$   &        \\
$^{	290	}$112&0.09  &1.87  &4.52  &3.89   &2.85   &2.99    & SF         &        \\
$^{	286	}$110&-1.44 &5.77  &8.10  &9.26   &6.99   &7.31    & SF         &        \\
$^{	282	}$108&-4.08 &8.36  &10.57 &12.71  &9.72   &10.26   & SF         &        \\
$^{	278	}$106&-7.64 &6.16  &8.15  &8.44   &6.39   &7.03    & SF         &        \\
\hline
$^{	320	}$126 &4.36 &-6.91 &-2.45 &-6.47  &-4.50 &-4.82  &$ \alpha $&         \\
$^{	316	}$124 &-2.58&4.98  &8.66  &12.40  &9.11  &8.68   &SF&         \\
$^{	312	}$122 &-1.25&-5.14 &-1.07 &-4.52  &-3.27 &-3.43  &$ \alpha $&         \\
$^{	308	}$120 &-1.01&5.18  &8.52  &11.47  &8.35  &8.25   &SF        &         \\
$^{	304	}$118 &-3.25&1.18  &4.41  &4.21   &2.93  &2.97   &SF        &         \\
$^{	300	}$116 &3.16 &0.68  &3.82  &2.91   &1.93  &2.11   &$ \alpha $&         \\
$^{	296	}$114 &5.15 &3.48  &6.33  &6.70   &4.72  &5.07   &$ \alpha $&         \\
$^{	292	}$112 &0.63 &2.59  &5.34  &4.71   &3.21  &3.68   &SF        &         \\
$^{	288	}$110 &-4.75&5.35  &7.85  &8.35   &5.94  &6.64   &SF        &         \\
$^{	284	}$108 &-4.20&8.52  &10.90 &12.76  &9.33  &10.31  &SF        &         \\
$^{	280	}$106 &-6.09&11.02 &13.42 &16.19  &12.03 &13.32  &SF        &         \\
\hline																	
\end{longtable}
\end{center}


\begin{center}
\small
\setlength{\tabcolsep}{3pt}
\begin{longtable}{|c|c|c|c|c|c|c||c|c|c|c|c|c|c|}
\caption{The predicted $\alpha$-decay half-lives of the Z $=$ 106, 114, 120, and 126 isotopes (even–even) with the MMF, MSF, MHF, SemFIS2, and UNIV formulas. The Q$_{\alpha}$ values are obtained by using RMF calculations using DD-ME2 parameter.}\label{hl}\\
\hline
 \multicolumn{1}{|c}{N}&
 \multicolumn{1}{|c}{Q$_{\alpha}(MeV)$ }&
 \multicolumn{5}{|c|}{$log_{10}T^{\alpha}_{1/2}(s)$} &

  \multicolumn{1}{|c}{N}&
 \multicolumn{1}{|c}{Q$_{\alpha}(MeV)$ }&
 \multicolumn{5}{|c|}{$log_{10}T^{\alpha}_{1/2}(s)$} \\
 \cline{3-7}
 \cline{10-14}
 \multicolumn{1}{|c}{}&
 \multicolumn{1}{|c}{${(DD-ME2)}$}&
 \multicolumn{1}{|c}{MMF}&
  \multicolumn{1}{|c}{MSF}&
   \multicolumn{1}{|c}{MHF}&
     \multicolumn{1}{|c}{SemFIS2}&
 \multicolumn{1}{|c|}{UNIV}&

 \multicolumn{1}{|c}{}&
 \multicolumn{1}{|c}{${(DD-ME2)}$}&
 \multicolumn{1}{|c}{MMF}&
  \multicolumn{1}{|c}{MSF}&
   \multicolumn{1}{|c}{MHF}&
     \multicolumn{1}{|c}{SemFIS2}&
 \multicolumn{1}{|c|}{UNIV}\\
\hline
\endfirsthead
\multicolumn{14}{c}%
{\tablename\ \thetable\ - \textit{Continued from previous page}} \\
   \hline
 \multicolumn{1}{|c}{N}&
 \multicolumn{1}{|c}{Q$_{\alpha}(MeV)$ }&
 \multicolumn{5}{|c|}{$log_{10}T^{\alpha}_{1/2}(s)$} &

 \multicolumn{1}{|c}{N}&
 \multicolumn{1}{|c}{Q$_{\alpha}(MeV)$ }&
 \multicolumn{5}{|c|}{$log_{10}T^{\alpha}_{1/2}(s)$} \\
 \cline{3-7}
 \cline{10-14}
 \multicolumn{1}{|c}{}&
 \multicolumn{1}{|c}{${(DD-ME2)}$}&
 \multicolumn{1}{|c}{MMF}&
  \multicolumn{1}{|c}{MSF}&
   \multicolumn{1}{|c}{MHF}&
     \multicolumn{1}{|c}{SemFIS2}&
 \multicolumn{1}{|c|}{UNIV}&

 \multicolumn{1}{|c}{}&
 \multicolumn{1}{|c}{${(DD-ME2)}$}&
 \multicolumn{1}{|c}{MMF}&
  \multicolumn{1}{|c}{MSF}&
   \multicolumn{1}{|c}{MHF}&
     \multicolumn{1}{|c}{SemFIS2}&
 \multicolumn{1}{|c|}{UNIV}\\
\hline
\endhead
\hline \multicolumn{14}{r}{\textit{Continued on next page}} \\
\endfoot
\hline
\endlastfoot
 \hline
\multicolumn{14}{|c|}{Z=106}\\
\hline
140	&11.25 & -7.40 &-4.51 &-5.69 &-6.09 &-5.26&178&5.54&12.59 &15.58 &18.61 &12.62  &15.26 \\
142	&10.93 & -6.57 &-3.95 &-4.92 &-5.20 &-4.60&180&6.65&8.13  &10.90 &10.48 &6.10   &8.87  \\
144	&10.52 & -5.50 &-3.17 &-3.90 &-4.09 &-3.71&182&5.14&14.57 &18.38 &22.03 &13.45  &17.96 \\
146	&10.11 & -4.37 &-2.31 &-2.78 &-2.91 &-2.73&184&4.86&16.00 &20.43 &24.83 &14.36  &20.14 \\
148	&9.85  & -3.60 &-1.70 &-2.04 &-2.08 &-2.09&186&8.12&3.28  &6.96  &1.83  &-1.30  &2.58  \\
150	&10.13 & -4.24 &-2.22 &-2.88 &-2.68 &-2.85&188&7.39&5.82  &9.58  &5.39  &0.28   &5.38  \\
152	&9.73  & -3.09 &-1.26 &-1.75 &-1.53 &-1.84&190&5.72&12.43 &16.72 &16.64 &6.07   &13.92 \\
154	&10.02 & -3.76 &-1.76 &-2.68 &-2.25 &-2.64&192&5.51&13.43 &18.13 &18.29 &5.91   &15.22 \\
156	&9.36  & -1.90 &-0.16 &-0.70 &-0.41 &-0.89&194&4.75&17.00 &22.90 &25.71 &8.62   &20.83 \\
158	&8.70  & 0.12  &1.64  &1.59  &1.65  &1.10 &196&3.17&24.84 &37.08 &49.55 &19.05  &38.73 \\
160	&8.17  & 1.82  &3.24  &3.60  &3.40  &2.83 &198&3.51&23.34 &33.78 &43.21 &13.71  &33.88 \\
162	&8.17  & 1.95  &3.46  &3.57  &3.33  &2.82 &200&3.39&23.98 &35.33 &45.28 &12.38  &35.36 \\
164	&8.31  & 1.59  &3.26  &2.88  &2.66  &2.28 &202&3.15&25.25 &38.56 &50.33 &11.99  &39.02 \\
166	&8.36  & 1.53  &3.34  &2.56  &2.29  &2.07 &204&2.80&26.81 &43.53 &58.54 &12.32  &44.96 \\
168	&7.81  & 3.38  &5.18  &4.86  &4.04  &4.02 &206&2.06&28.14 &57.21 &82.52 &17.18  &62.36 \\
170	&7.58  & 4.27  &6.17  &5.88  &4.65  &4.91 &208&1.50&23.72 &73.72 &111.95&21.76  &83.58 \\
172	&7.07  & 6.16  &8.15  &8.44  &6.39  &7.03 &210&1.49&23.64 &74.52 &113.09&16.67  &84.04 \\
174 &5.86  &11.02  &13.42 &16.19 &12.03 &13.32&212&1.21&15.78 &87.19 &136.00&16.07  &100.19\\
176 &5.90  &10.95  &13.51 &15.83 &11.19 &13.03&   &    &      &      &      &       &      \\
\hline                                                                 \multicolumn{14}{|c|}{Z=114}\\
\hline
160&11.63 &-6.33 &-3.26 &-4.72 &-3.76 &-4.15 &  192 &8.52&3.21  &6.94    &4.83   &1.33  &4.03   \\
162&11.32 &-5.54 &-2.60 &-3.90 &-2.97 &-3.49 &  194 &5.44&14.55 &19.65   &26.06  &14.29 &19.75  \\
164&11.60 &-6.10 &-3.04 &-4.77 &-3.59 &-4.16 &  196 &4.25&20.06 &28.08   &40.15  &21.98 &30.24  \\
166&11.77 &-6.38 &-3.23 &-5.32 &-3.96 &-4.56 &  198 &7.29&7.51  &11.85   &11.27  &3.30  &8.95   \\
168&11.51 &-5.70 &-2.61 &-4.68 &-3.38 &-4.01 &  200 &7.22&7.83  &12.41   &11.55  &2.60  &9.23   \\
170&11.20 &-4.91 &-1.89 &-3.90 &-2.73 &-3.35 &  202 &7.07&8.48  &13.33   &12.44  &2.15  &9.94   \\
172&10.97 &-4.28 &-1.27 &-3.32 &-2.26 &-2.84 &  204 &6.21&11.82 &17.28   &18.73  &4.32  &14.51  \\
174&9.11  &0.65  &3.16  &3.31  &2.95  &2.31  &  206 &6.08&12.45 &18.25   &19.73  &3.54  &15.28  \\
176&9.27  &0.28  &2.93  &2.55  &2.22  &1.76  &  208 &5.86&13.43 &19.68   &21.56  &2.99  &16.63  \\
178&9.47  &-0.20 &2.62  &1.60  &1.33  &1.09  &  210 &5.57&14.76 &21.58   &24.25  &2.62  &18.59  \\
180&9.04  &1.10  &3.93  &3.27  &2.41  &2.42  &  212 &5.27&16.12 &23.62   &27.21  &2.11  &20.72  \\
182&8.28  &3.48  &6.33  &6.70  &4.72  &5.07  &  214 &5.01&17.39 &25.62   &30.10  &1.30  &22.79  \\
184&8.22  &3.75  &6.74  &6.88  &4.50  &5.25  &  216 &4.79&18.47 &27.43   &32.68  &0.10  &24.63  \\
186&10.41 &-2.29 &1.40  &-2.59 &-2.54 &-1.67 &  218 &4.59&19.50 &29.26   &35.29  &-1.36 &26.47  \\
188&9.77  &-0.53 &3.12  &-0.35 &-1.28 &0.06  &  220 &4.40&20.49 &31.08   &37.92  &-3.10 &28.32  \\
190&9.13  &1.31  &4.97  &2.13  &0.02  &1.97  &      &    &      &        &       &      &       \\
\hline
\multicolumn{14}{|c|}{Z=120}\\
\hline
160	&13.42 	&-9.27 &-5.09  &-7.59 &-6.19  &-6.11   & 188  & 8.03  & 5.18    & 8.52   &11.47   &8.35   &8.25       \\  162	&12.77 	&-7.92 &-4.04  &-6.00 &-4.79  &-4.91   & 190  & 6.80  & 9.53    & 13.41  &19.38   &13.77  &14.15      \\
164	&13.32 	&-8.92 &-4.92  &-7.36 &-5.71  &-5.99   & 192  & 5.92  & 13.00   & 17.78  &26.48   &18.35  &19.43      \\
166	&13.28 	&-8.76 &-4.83  &-7.30 &-5.53  &-5.95   & 194  & 6.32  & 11.51   & 16.02  &23.08   &15.16  &16.85      \\
168	&12.94 	&-8.01 &-4.22  &-6.49 &-4.78  &-5.34   & 196  & 7.65  & 6.80    & 10.76  &13.50   &7.83   &9.81       \\
170	&11.66 	&-5.24 &-1.89  &-2.99 &-1.93  &-2.68   & 198  & 6.83  & 9.77    & 14.23  &19.03   &10.85  &13.85      \\
172	&10.51 	&-2.47 &0.57   &0.73  &1.11   &0.17    & 200  & 8.61  & 3.93    & 8.13   &7.77    &2.86   &5.79       \\
174	&12.29 	&-6.45 &-2.87  &-4.96 &-3.37  &-4.14   & 202  & 8.29  & 5.01    & 9.39   &9.39    &3.23   &7.00       \\
176	&11.97 	&-5.67 &-2.16  &-4.10 &-2.67  &-3.47   & 204  & 7.98  & 6.08    & 10.69  &11.06   &3.49   &8.24       \\
178	&11.80 	&-5.23 &-1.72  &-3.70 &-2.36  &-3.14   & 206  & 7.71  & 7.03    & 11.88  &12.56   &3.53   &9.36       \\
180	&10.86 	&-2.99 &0.30   &-0.73 &-0.11  &-0.90   & 208  & 8.44  & 4.81    & 9.78   &8.08    &0.22   &6.31       \\
182	&9.74  	&-0.08 &3.03   &3.42  &2.99   &2.21    & 210  & 8.67  & 4.18    & 9.36   &6.60    &-1.37  &5.38       \\
184	&9.69  	&0.15  &3.34   &3.57  &2.94   &2.35    & 212  & 6.89  & 10.17   & 15.97  &17.93   &3.25   &13.30      \\
186	&9.72  	&0.17  &3.47   &3.37  &2.59   &2.23    & 214  & 6.81  & 10.55   & 16.63  &18.43   &2.24   &13.70      \\
\hline
\multicolumn{14}{|c|}{Z=126}\\
\hline
164	&16.84 &-14.13  &-8.49  &-13.36  &-10.27  &-10.04  &194 &13.26 &-6.91 &-2.45  &-6.47  &-4.50  &-4.82          \\ 166	&15.78 &-12.43  &-7.24  &-11.36  &-8.68   &-8.62   &196 &10.20 &-0.04 &3.88   &4.05   &2.73   &2.57           \\
168	&14.60 &-10.40  &-5.67  &-8.86   &-6.69   &-6.82   &198 &10.17 &0.12  &4.14   &4.08   &2.45   &2.63           \\
170	&13.54 &-8.40   &-4.07  &-6.31   &-4.63   &-4.97   &200 &12.85 &-5.87 &-1.18  &-5.77  &-4.50  &-4.10          \\
172	&13.61 &-8.47   &-4.18  &-6.50   &-4.65   &-5.15   &202 &12.70 &-5.48 &-0.70  &-5.48  &-4.54  &-3.81          \\
174	&14.73 &-10.40  &-5.82  &-9.21   &-6.58   &-7.14   &204 &12.46 &-4.91 &-0.06  &-4.91  &-4.45  &-3.34          \\
176	&14.83 &-10.49  &-5.91  &-9.49   &-6.69   &-7.33   &206 &12.18 &-4.22 &0.70   &-4.15  &-4.29  &-2.74          \\
178	&14.83 &-10.40  &-5.84  &-9.55   &-6.65   &-7.35   &208 &11.59 &-2.85 &2.10   &-2.26  &-3.53  &-1.38          \\
180	&16.06 &-12.33  &-7.39  &-12.31  &-8.60   &-9.23   &210 &11.11 &-1.68 &3.34   &-0.66  &-3.03  &-0.21          \\
182	&14.62 &-9.89   &-5.38  &-9.25   &-6.31   &-7.08   &212 &10.89 &-1.06 &4.09   &0.05   &-3.14  &0.36           \\
184	&14.31 &-9.26   &-4.82  &-8.60   &-5.81   &-6.58   &214 &10.30 &0.47  &5.71   &2.36   &-2.48  &2.00           \\
186	&14.21 &-9.00   &-4.55  &-8.47   &-5.70   &-6.44   &216 &9.91  &1.56  &6.95   &3.96   &-2.32  &3.16           \\
188	&12.70 &-6.04   &-1.96  &-4.45   &-2.79   &-3.59   &218 &9.55  &2.60  &8.17   &5.51   &-2.32  &4.29           \\
190	&13.49 &-7.53   &-3.16  &-6.85   &-4.59   &-5.21   &220 &9.25  &3.50  &9.28   &6.86   &-2.54  &5.28           \\
192	&13.46 &-7.38   &-2.95  &-6.90   &-4.70   &-5.18   &    &      &      &       &       &       &               \\
\end{longtable}
\end{center}
\end{document}